%% file: main.tex
\documentclass[prx,aps,longbibliography,showpacs,groupedaddress,superscriptaddress,twocolumn,toc=flat,nofootinbib]{revtex4-1}
\input{header}

\usepackage{graphicx}
\usepackage[justification=justified,singlelinecheck=false]{caption} 
\usepackage[unicode=true,
 bookmarks=false,
 breaklinks=false,pdfborder={0 0 1},backref=false,colorlinks=true,]
 {hyperref}
\hypersetup{
 linkcolor=[rgb]{0,0,1},citecolor=[rgb]{0,0,1},urlcolor=[rgb]{0,0,1}}

\AtBeginDocument{%
    \newwrite\bibnotes
    \def\bibnotesext{Notes.bib}
    \immediate\openout\bibnotes=\jobname\bibnotesext
    \immediate\write\bibnotes{@CONTROL{REVTEX41Control}}
    \immediate\write\bibnotes{@CONTROL{%
    apsrev41Control,author="08",editor="1",pages="1",title="0",year="1"}}
     \if@filesw
     \immediate\write\@auxout{\string\citation{apsrev41Control}}%
    \fi
}%

\begin{document}
\title{Topology meets symmetry breaking: Hidden order, intrinsically gapless topological states and finite-temperature topological transitions}

\author{Reja H. Wilke}
\affiliation{Department of Physics and Arnold Sommerfeld Center for Theoretical Physics (ASC), Ludwig-Maximilians-Universit\"at M\"unchen, Theresienstr. 37, M\"unchen D-80333, Germany}
\affiliation{Munich Center for Quantum Science and Technology (MCQST), Schellingstr. 4, M\"unchen D-80799, Germany}
\affiliation{Institute for Theoretical Physics, ETH Z\"urich, CH-8093 Zurich, Switzerland}

\author{Henning Schl\"omer}
\affiliation{Department of Physics and Arnold Sommerfeld Center for Theoretical Physics (ASC), Ludwig-Maximilians-Universit\"at M\"unchen, Theresienstr. 37, M\"unchen D-80333, Germany}
\affiliation{Munich Center for Quantum Science and Technology (MCQST), Schellingstr. 4, M\"unchen D-80799, Germany}

\author{Simon M. Linsel}
\affiliation{Department of Physics and Arnold Sommerfeld Center for Theoretical Physics (ASC), Ludwig-Maximilians-Universit\"at M\"unchen, Theresienstr. 37, M\"unchen D-80333, Germany}
\affiliation{Munich Center for Quantum Science and Technology (MCQST), Schellingstr. 4, M\"unchen D-80799, Germany}

\author{Annabelle Bohrdt}
\affiliation{Department of Physics and Arnold Sommerfeld Center for Theoretical Physics (ASC), Ludwig-Maximilians-Universit\"at M\"unchen, Theresienstr. 37, M\"unchen D-80333, Germany}
\affiliation{Munich Center for Quantum Science and Technology (MCQST), Schellingstr. 4, M\"unchen D-80799, Germany}

\author{Fabian Grusdt}
\affiliation{Department of Physics and Arnold Sommerfeld Center for Theoretical Physics (ASC), Ludwig-Maximilians-Universit\"at M\"unchen, Theresienstr. 37, M\"unchen D-80333, Germany}
\affiliation{Munich Center for Quantum Science and Technology (MCQST), Schellingstr. 4, M\"unchen D-80799, Germany}

\date{\today}

\begin{abstract}
Since the discovery of phase transitions driven by topological defects, the classification of phases of matter has been significantly extended beyond Ginzburg and Landau's paradigm of spontaneous symmetry breaking (SSB). In particular, intrinsic and symmetry-protected topological (SPT) orders have been discovered in (mostly gapped) quantum many-body ground states. However, these are commonly viewed as zero-temperature phenomena, and their robustness in a gapless ground state or against thermal fluctuations remains challenging to tackle. Here we introduce an explicit construction for SPT-type states with hidden order associated with SSB: They feature (quasi) long-range correlations along appropriate edges, but short-range order in the bulk; ground state degeneracy associated with SSB; and non-local string order in the bulk. We apply our construction to predict two types of finite-temperature SPT transitions protected by 1-form symmetries, in the Ising and BKT class respectively, where the usual signs of criticality appear despite the absence of a diverging correlation length in the bulk. While the state featuring hidden Ising order is gapped, the other SPT state associated with the BKT-SPT transition has hidden $U(1)$, or XY-order and constitutes an intrinsically gapless SPT state, associated with a gapless Goldstone mode. Specifically, in this work we discuss spins with global $\mathbb{Z}_2$ or $U(1)$ symmetry coupled to link variables constituting a loop gas model with a 1-form symmetry. By mapping this system to an Ising-gauge theory, we demonstrate that one of the SPT phases we construct corresponds to the Higgs-SPT phase at $T=0$ -- which we show here to remain stable at finite temperature. Our work paves the way for a more systematic search for hidden order SPT phases, including in gapless systems, and raises the question if a natural (finite-$T$) spin liquid candidate exists that realizes hidden order in the Higgs-SPT class.
\end{abstract}
\maketitle

The Ginzburg-Landau paradigm connects long-range order (LRO) to the spontaneous breaking of an underlying symmetry. Together with the subsequently discovered notion of topological phase transitions, driven by the proliferation of topological defects~\cite{berezinskii1972,Kosterlitz1973,Levin2005}, this forms the backbone of today's classification of the phases of matter. While spontaneously symmetry-broken (SSB) orders are thoroughly understood~\cite{Sachdev2011}, both at temperatures $T=0$ and $T>0$, the complete characterization of topological orders, including symmetry-protected topological (SPT) states, remains an ongoing task. Although topology in gapped one-dimensional systems is well classified~\cite{Chen2011,Pollmann2012}, higher-dimensional settings~\cite{KITAEV2003,Kitaev2006a}, gapless topological phases~\cite{Scaffidi2017,Thorngren2020,Verresen2021a} as well as interacting systems~\cite{Chen2014a,Mai2023} remain subject of ongoing research, to name a few. 

Another avenue of active research concerns the fate of topological states of matter at finite temperatures, $T>0$. While intrinsic topological order is not robust to thermal fluctuations in two spatial dimensions, it can survive e.g. in the form of a passively protected quantum memory in four dimensions~\cite{Dennis2002,Alicki2010}. Since two-point correlations in the bulk decay exponentially in topological phases, most characterizations of topological order, intrinsic or SPT, rely on the analysis of the non-local structure of the entanglement~\cite{Kitaev2006,Li2008,Chen2010}, which becomes cumbersome at finite temperatures~\cite{Grusdt2017LU}. Nevertheless, generalizations of topological invariants to mixed states~\cite{Diehl2011,Uhlmann1986,Rivas2013a,Budich2015a,Bardyn2018, Huang2024, Huang2025,Huang2024coherentinformationmixedstatetopological,Ma2023,Ma2025, Zhang2024} and finite temperatures~\cite{Huang2014,Molignini2023, HuangPRL2025, Roberts2017} have been proposed. While a no-go theorem rules out the existence of entangled SPT states protected by a global 0-form symmetry~\cite{Roberts2017} at $T>0$, a general understanding when and which features of SPT states can remain stable at finite temperatures remains lacking.

In this article, we construct a class of SPT states featuring hidden order (HO), or hidden SSB (hSSB). While the connection of SSB and SPT~\cite{Kennedy1992} as well as hidden order and SPT~\cite{Else2013, Levin2012,Hung2012, Thorngren2020} have been previously studied at $T=0$, here we show that such HO-SPT states in general remain robust at finite temperatures in $d \geq 2$ dimensions, and constitute intrinsically gapless SPT phases when the underlying protecting, global symmetry is continuous. The basic strategy is related to the construction of SPT phases from decorated domain walls~\cite{Chen2014a,Parker2018}, which naturally leads us to the Higgs-SPT phase recently discovered in the $\mathbb{Z}_2$ Ising gauge theory (IGT)~\cite{Verresen2024,Hiroki2024,Rakovszky2023}. Similarly, antiferromagnetic (AFM) order hidden by fluctuating stripes was recently proposed as a description of the pseudogap phase of the cuprate superconductors~\cite{Schloemer2024}, see also~\cite{Zhang2002,Podolsky2005}. In this article we clarify how all these scenarios can be viewed as forms of HO-SPT phases, which remain robust at finite temperature and naturally extend to intrinsically gapless SPT states. 

\begin{figure*}
	\includegraphics[width=\textwidth]{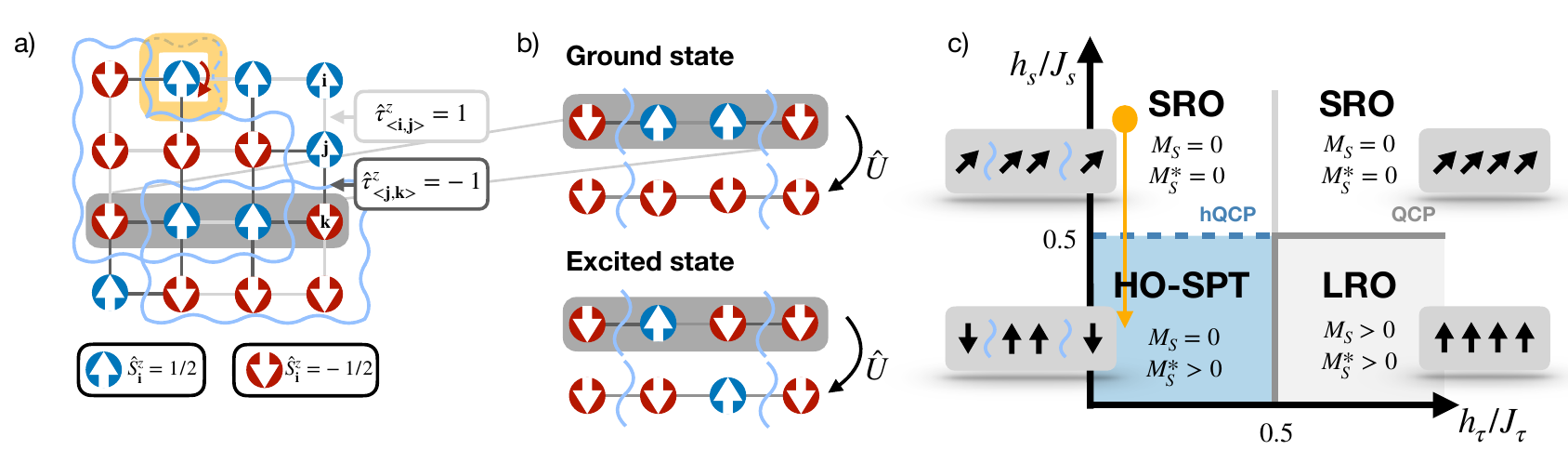}
	\caption{\justifying We predict a class of HO-SPT phases featuring SSB without a local bulk order parameter, in systems where domain walls of a spin order parameter couple to a loop gas of fluctuating link variables. a) In the hidden-Ising order (HIO) model, the sign of Ising interactions $\propto \hat{S}^z_{\vec{i}} \hat{S}^z_{\vec{j}}$ between spins is flipped across strings defined by link variables $\tau^z_{\nn}=-1$ (blue, wiggly lines). Fluctuations of the strings around sites (highlighted in yellow) are accompanied by flips of the central spin. In addition, spin-flips driven by a transverse field can be included. The 1D HIO model can be understood as an excerpt of the 2D model (dark gray box).
    b) Deep in the HO-SPT phase, the ground state satisfies the hidden order rule, illustrated in the top panel: spins flip sign only across a flipped link. The bottom panel shows an excitation, corresponding to a spin-flip without a link-flip. Spins and link variables in the HIO model at $\lambda=0$ can be exactly decoupled by the non-local unitary transformation $\hat{U}$ illustrated in both panels, which flips all spins between two strings with $\tau^z=-1$.
    c) The phase diagrams of the HIO model in 1D, Eq.~\eqref{eq:1d_ham} for $\lambda = 0$, and in 2D, Eq.~\eqref{eq:TC_meets_Ising}, have the same structure [only the critical values in 2D, $(h_\tau/J_\tau)_{\rm c,2D}$ and $(h_S/J_S)_{\rm c,2D}$, differ from their shown value $0.5$ in 1D]. The conventional SSB phase with bulk long-range order (LRO) and non-zero magnetization $M_S>0$ is located next to a symmetric phase with short-range order (SRO) and the SPT phase with hidden order (HO-SPT), which features hSSB. As explained in the text, the HO can be detected by the magnetization of Ising spins in squeezed space, $M_S^*$. A hidden quantum critical point (hQCP), or HO-SPT transition, separates the HO-SPT from the SRO phase. The orange arrow in c) indicates the corresponding scan in our numerics shown in Fig.~\ref{fig:pd_lamda} a).
    }
	\label{fig:1d_pd}
\end{figure*}

Our starting point is a model of spins, locally coupled to fluctuating link variables, which has a global symmetry $\hat{\mathcal{S}}$ associated with the spins, i.e. $[\hat{H},\hat{\mathcal{S}}]=0$ where $\hat{H}$ is the Hamiltonian. We further assume that the link variables form a loop gas model~\cite{Levin2005}, constituted by closed string configurations on the dual lattice, see Fig.~\ref{fig:1d_pd} a) for an illustration. When the underlying spin-link interactions lead to binding of domain walls of the local spin-order parameter to the strings of the loop gas, sufficiently strong fluctuations of the latter can lead to short-range correlations between spins in the bulk. Nevertheless, the underlying global symmetry $\hat{\mathcal{S}}$ of the spins can remain spontaneously broken, turning the expected long-range correlations associated with this SSB into non-local string order in the bulk. By this construction, an SPT phase protected by $\hat{\mathcal{S}}$ is obtained.  Crucially, this construction does not rely on the topological ground-state of the loop gas model.

To describe HO-SPT phases featuring such hSSB, we provide explicit solutions of different microscopic models in this class. To this end, we construct an exact, non-local unitary transformation $\hat{U}$ that allows to decouple the original Hamiltonian into two independent parts: a conventional spin model, symmetric under $\hat{\mathcal{S}}$, and a fluctuating loop gas model realized as a perturbed toric code in a field~\cite{Fradkin1979,Trebst2007,Wu2012b}. The motivation for this transformation derives from the concept of squeezed space originally introduced to describe doped AFMs~\cite{OgataShiba,Kruis2004,Schloemer2024} and applied for the construction of intrinsically gapless, one-dimensional (1D) SPT states~\cite{Thorngren2020}. As a consequence of the strong correlations between spins and links, the unitary $\hat{U}$ basically unwinds the domain walls of the spin order parameter and turns the hidden order into conventional long-range order of the spins in the newly constructed basis, see Fig.~\ref{fig:1d_pd} b) for an illustration. 

This construction will lead us to the generic HO phase diagram shown in Fig.~\ref{fig:1d_pd} c), where the HO-SPT phase is located next to a conventional, ordered phase and a trivial, symmetric phase. The first transition (HO-SPT to LRO) appears to be conventional symmetry-breaking in the bulk, although it is driven by the confinement of the loop gas and hence in a different universality class in general. The second transition (HO-SPT to SRO) is of SPT type and constitutes a hidden quantum critical point (hQCP): Since both sides of the transition have short-range correlations in the bulk, it cannot be characterized by a diverging correlation length. However, the correlation length associated with the hidden, non-local string order diverges, and likewise other characteristics of quantum criticality, such as critical transport or collective mode softening, remain present. 

In principle, our construction can be adapted to any symmetry $\hat{\mathcal{S}}$. In this paper, we explicitly consider the discrete $\mathbb{Z}_2$ and the continuous $U(1)$ symmetries. One of the most striking consequences of the hSSB in the HO-SPT phase is its robustness at finite temperatures, $T>0$ in two or more dimensions: Since the hidden order can be described by a conventional spin system decoupled from the loop gas, it inherits the usual Ginzburg-Landau classification of SSB. Indeed, the underlying symmetry $\hat{\mathcal{S}}$ is broken globally, rather than locally, while LRO is hidden, rather than destroyed, by the fluctuating loop gas.  In this paper we explicitly discuss hidden Ising and BKT orders, associated with $\mathbb{Z}_2$ and $U(1)$ symmetries. 

Our paper is organized as follows: In the first section, we illustrate the connection between SSB, SPT and HO for the example of two coupled 1D transverse-field Ising models (TFIMs). An almost identical model was recently discussed~\cite{Verresen2024}, and our results completely agree with their conclusions derived from considering the spontaneous breaking of higher-form symmetries~\cite{Wen2019}. We provide a new perspective by explicitly constructing the HO unitary transformation $\hat{U}$, see Fig.~\ref{fig:1d_pd} b), which decouples the \emph{entire} spectrum of the Hamiltonian. This paves the way for our subsequent extensions to higher dimensions, continuous symmetries and finite temperature.

In the second section, we discuss hidden order associated with a discrete $\mathbb{Z}_2$ symmetry in 2D, where we consider a transverse-field Ising model (TFIM) of spins coupled to a perturbed toric code. By a mapping to the double-Higgs IGT, we argue that the HO-SPT phase we identify coincides with the Higgs-SPT phase of the $\mathbb{Z}_2$ IGT~\cite{Verresen2024,Xu2025}. As a central new result, we demonstrate that the HO-SPT phase is robust to thermal fluctuations and gives rise to a finite-$T$ SPT transition.

In the third section, we extend our results to hidden order associated with a continuous $U(1)$ symmetry in 2D. In the ground state, we find an intrinsically gapless HO-SPT phase with hidden $U(1)$ or XY order and a gapless Goldstone mode. This phase essentially survives at finite temperatures, although with hidden quasi-long range order and power-law correlations at the edge, before it disappears in a finite-$T$ SPT transition of BKT type. We predict these physics in an IGT coupled to a $U(1)$ matter field, closely related to the classical 3D XY model~\cite{Sachdev2019}.

Our paper closes with an outlook and a discussion how HO-SPT phases may be experimentally observed, in real materials and synthetic quantum matter.

% % % % % % % % % % % % % % % % % % % % % % % % % % % % 
\section*{Hidden Order in 1D: SPT $=$ hidden SSB}
\label{sec:1d_HO}
% % % % % % % % % % % % % % % % % % % % % % % % % % % % 
We start by explaining the fundamental idea how hSSB and HO can be realized in an exactly solvable model in one dimension (1D). We provide an explicit construction of a (HO-) SPT phase in 1D, which by itself is well understood~\cite{Briegel2001,Verresen2024}. However, as we show below, the construction we make can be straightforwardly generalized to higher dimensions, finite temperature or continuous symmetries and provides valuable insights into the relation of SPT and SSB orders.

We consider a 1D lattice with spin-$1/2$ degrees of freedom residing both on the lattice sites $j$ and on the links $\langle j,j+1 \rangle$, see Fig.~\ref{fig:1d_pd} a) and b). We define the following Hamiltonian, which we refer to as the 1D hidden-Ising order (HIO) Hamiltonian,
\begin{equation}
\begin{aligned}
	\hat H &= - J_S \sum_{j} \hat S_{j+1}^z \hat S_j^z [\hat \tau^z_{\langle j, j+1\rangle}(1-\lambda) + \lambda]
    + h_S \sum_j \hat S^x_j\\
    &- h_\tau \sum_j \hat \tau^z_{\langle j,j+1\rangle}
    + J_\tau \sum_j \hat \tau^x_{\langle j-1,j\rangle} \hat\tau^x_{\langle j,j+1\rangle} \hat S^x_{j}
    \;.
    \label{eq:1d_ham}
\end{aligned}
\end{equation}
Here, $\hat S^\alpha_j$ denotes the $\alpha$-component of the spin on site $j$ with $\alpha = x,y,z$, and $\hat \tau^\alpha_{\langle j,j+1 \rangle}$ refers to Pauli matrices on the links between neighboring sites.
The model exhibits a NN Ising-like interaction $\propto J_S$ and a transverse field $h_S$.
The real parameter $\lambda \in [0,1]$ interpolates between a conventional NN Ising interaction (for $\lambda=1$) and one with a sign-flip controlled by the $\hat \tau^z$-field on the link connecting both sites (for $\lambda=0$). A very similar model was constructed starting from the cluster model ($h_S=h_\tau=\lambda=0$) to construct the same SPT phase~\cite{Verresen2024} that we will discuss now.

In the following we focus on the ordered and disordered phases of the spins $\hat{\vec{S}}_{j}$, associated with the global $\mathbb{Z}_2$ symmetry $\hat{S}^z_j \to - \hat{S}^z_j$ of the HIO model, Eq.~\eqref{eq:1d_ham}. The model has an additional, global $\mathbb{Z}_2$ symmetry, $\hat{\tau}^x_j \to - \hat{\tau}^x_j$, which is not important for the following discussion and can be broken by a weak longitudinal field term, $b_\tau \sum_j \hat{\tau}^x_{\langle j,j+1 \rangle}$, without changing the nature of the observed phase transitions of the spins $\hat{\vec{S}}_{j}$. Moreover, the HIO model at $b_\tau=\lambda=0$ has a self-duality $\tau^z \leftrightarrow \hat{S}^x$ and a local $\mathbb{Z}_2$ gauge symmetry~\cite{Borla2021,Kebric2024}, neither of which will be essential for the physics that we describe now.

When the link field $\hat \tau^z = 1$ is fully polarized, for large $h_\tau \gg J_\tau \geq 0$, the 1D HIO Hamiltonian reduces to a TFIM. In this limit, it exhibits a well-known quantum critical point (QCP) describing a transition from a SSB ferromagnet (FM) to a paramagnet (PM)~\cite{Sachdev2011,Simon2011}, assuming $J_S>0$ and tuning $h_S$. The situation becomes more interesting when quantum fluctuations of the link variables, introduced by the last term $\propto J_\tau$ in Eq.~\eqref{eq:1d_ham}, dominate, $J_\tau \gg h_\tau \geq 0$: On one hand, this tends to de-polarize the links $\hat \tau^z$; on the other hand, the operator $\hat{S}^x_j$ appearing in this term introduces spin flips in the FM. By construction of the HIO Hamiltonian, the latter lead to no additional energy cost $\propto J_S$ for $\lambda=0$ since the values of $\hat{\tau}^z_{\langle j, j\pm 1 \rangle}$ are also flipped by the term $\propto J_\tau$. I.e., by adding $J_\tau$ processes, we introduce fluctuating domain walls in the FM which are bound to $\hat \tau^z$ excitations -- as we show next, these can hide the FM order when they proliferate for small values of $h_\tau$, see Fig.~\ref{fig:1d_pd} c).

Now we construct an exact, non-local unitary transformation to demonstrate that the fluctuating domain walls destroy the long-range FM order of the spins $\hat{S}^z_j$ when $J_\tau$ and $J_S$ are sufficiently large. Moreover, this transformation allows for an exact solution of Eq.~\eqref{eq:1d_ham} in the case $\lambda=0$, and reveals that SSB still takes place even in the absence of long-range FM spin correlations. Our construction is similar to the idea of squeezed space~\cite{Kruis2004, HilkerScience2017}, introduced to solve the 1D $t-J$ model~\cite{OgataShiba}, and leads to a non-local string order parameter characterizing the hSSB phase of the 1D HIO model. Notably, the same construction will allow us to derive a similar solution of the HIO model in 2D, which we discuss in the next section.

The unitary transformation $\hat{U}$ we construct disentangles the spin configurations $\hat{S}^z_j=\pm 1/2$ from the fluctuating link variables $\hat \tau^z_{\langle j, j+1\rangle}=\pm 1$.
%, which we label by $\hat\tau^z_j$ in the following. 
We define $\hat{U}$ by its action on basis states $\ket{\{ S^z_j, \tau^z_{\langle j, j+1\rangle}\}}$ as follows,
\begin{equation}
    \hat U \ket{\{ S^z_j, \tau^z_{\langle j, j+1\rangle}\}} = \ket{\{ \tilde S^z_j = (-1)^{p_j} S^z_j, \tau^z_{\langle j, j+1\rangle} \}}\;,
    \label{eq:1d_HO_unitary}
\end{equation}
where $p_j$ denotes the number of negative links for $i<j$, i.e. $(-1)^{\hat{p}_j} = \prod_{i<j}\hat{\tau}^z_{\langle i, i+1\rangle}$.
The action of the unitary $\hat{U}$ is illustrated in Fig.~\ref{fig:1d_pd} b): A spin on site $j$ is flipped if and only if the number of negative links on sites $i<j$ is odd. Our transformation $\hat{U}$ can be viewed as a generalization of the unitary proposed in Ref.~\cite{Verresen2024} in the limit $J_\tau \to \infty$ for the ground state.

Applying the unitary on the Hamiltonian we obtain a representation of the 1D HIO model in the new basis, which we refer to as the \emph{squeezed space}, in analogy with doped 1D quantum magnets. The result are two decoupled TFIMs for spin $(S)$ and link $(\tau)$ degrees of freedom,
\begin{equation}
    \hat U^\dagger\hat H(\lambda=0)\hat U =\hat H^S_{\rm TFIM} + \hat H_{\rm TFIM}^\tau\;,
    \label{eq:decoupling_1d}
\end{equation}
see Appendix~\ref{sec:appA} for details. Hence, in squeezed space, the eigenstates of Eq.~\eqref{eq:decoupling_1d} at $\lambda=0$ factorize, $\ket{\psi} = \ket{\psi}_{S} \otimes \ket{\psi}_{\tau}$; here $\ket{\psi}_{S/\tau}$ denote eigenstates of the TFIM on sites/links. Since the latter can be expressed analytically by a combination of a Jordan-Wigner transformation and a Bogoliubov transformation~\cite{Coleman2015}, the 1D HIO model is integrable at $\lambda=0$. 

From Eq.~\eqref{eq:decoupling_1d}, we directly obtain the phase diagram of the HIO model at $\lambda = 0$,  shown in Fig.~\ref{fig:1d_pd}~c). It consists of two independent Ising-type phase transitions, of links and spins respectively, manifesting in crossing, straight lines in Fig.~\ref{fig:1d_pd} c). In squeezed space, the usual order parameters of the TFIM can be used to characterize the SSB Ising transition of the spins $\hat{S}^z$, such as magnetization $M^*_{S} = \langle |\frac{1}{L}\sum_j \hat S^z_j|\rangle_{\rm sq}$ or long-range spin-spin correlations $C^*(d) = \langle  \hat{S}^z_0 \hat{S}^z_d \rangle_{\rm sq}$. Here $\langle \cdot \rangle_{\rm sq} = \langle \hat{U} \cdot \hat{U}^\dagger \rangle$ is the expectation value in the new basis after applying $\hat{U}$, which turns the squeezed space order parameters into non-local string operators in the original model Eq.~\eqref{eq:1d_ham}, e.g. $C^*(d) = \langle  \hat{S}^z_0 \left( \prod_{0\leq j<d} \hat{\tau}^z_{\langle j, j+1\rangle} \right) \hat{S}^z_d \rangle$. In particular, the string order parameter $C^*(d)$ can retain long-range correlations when $\hat{\tau}^z$ fluctuates strongly, while the two-point spin-correlations $C(d) = \langle \hat{S}^z_0 \hat{S}^z_d \rangle \simeq e^{-d/\xi}$ decay exponentially in this regime, with $\xi$ the correlation length of the link variables: This leads to a hidden order, or SPT, phase which exhibits SSB without long-range correlations.

The unitary $\hat{U}$ also affects the link variables $\hat{\tau}^x_{\langle j, j+1 \rangle}$, by attaching a string of $\hat{S}^x$ operators, see Appendix~\ref{sec:appA}. Since our focus is on the physics of the spins $\hat{\vec{S}}_j$, in the following we will only consider order parameters for $\hat{\tau}$ variables in squeezed space, e.g. the link magnetization $M_\tau^* = \langle |\frac{1}{L}\sum_j \hat \tau^x_{\langle j, j+1\rangle}|\rangle_{\rm sq}$. Combining all order parameters, for spins and links, we arrive at the following groundstate phase diagram of the 1D HIO model at $\lambda=0$ which depends only on the ratios $h_\tau/J_\tau$ and $h_s/J_s$, cf. Fig.~\ref{fig:1d_pd}~c):

\begin{figure}
	\includegraphics[width=0.5\textwidth]{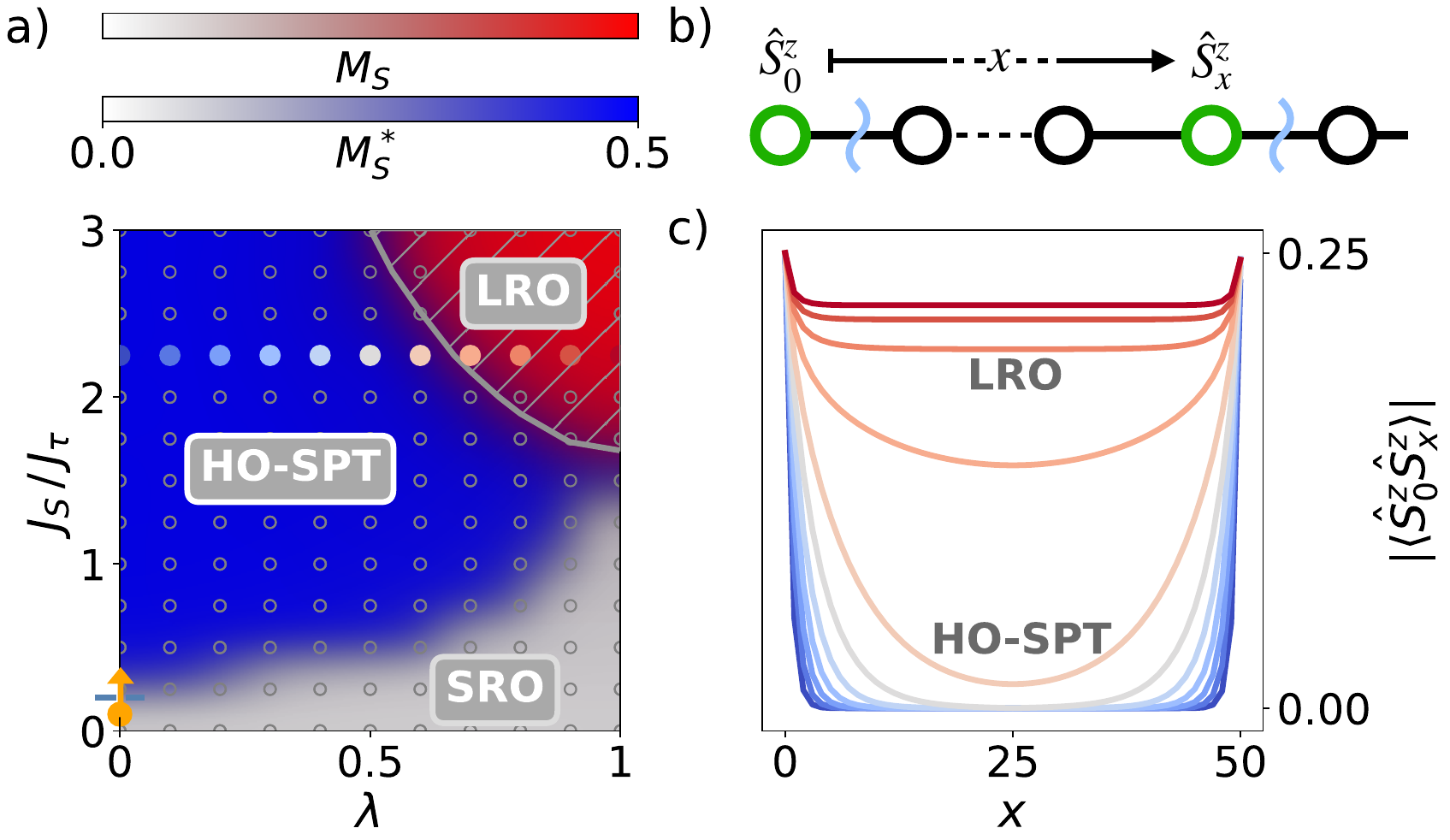}
	\caption{\justifying
    Numerical study of the HO-SPT phase in the 1D HIO model. 
    a) We perform DMRG simulations using the \textsc{SyTen} toolkit and take snapshots of the many-body wavefunction via the perfect sampling approach~\cite{Ferris2012,Buser2022}, allowing us to evaluate magnetizations $M_S$ and $M_S^*$ in real and squeezed space; the latter are indicated by two overlaid color maps. For the exactly solvable case $\lambda = 0$, the hQCP found in Fig.~\ref{fig:1d_pd} c) is located at a critical value $(J_S/J_\tau)_c = 0.2$, indicated by a short blue line. The orange arrow corresponds to the same scan in Fig.~\ref{fig:1d_pd} c). We find that for non-zero values of $\lambda$, the HO phase persists until eventually it transitions to the LRO phase (with $M_S>0$, for large $J_S/J_\tau$) or the disordered phase (with $M_S=M_S^*=0$, for small $J_S/J_\tau$). Areas with $\frac{1}{(L-1)}\langle | \sum_j \hat{\tau}^z_{\langle j,j+1 \rangle} | \rangle > 0.3$ are indicated by hatched, grey lines. b) We define standard spin-spin correlations $\langle |\hat S^z_0 \hat S^z_x|\rangle$ as a function of distance $x$, computed along the transition from HO to LRO at $J_S/J_\tau=2.25$ in c). 
    Colors in c) correspond to different values of $\lambda$, as highlighted by data points of the same color in a). We find that the bulk order disappears around $\lambda = 0.7$, whereas long-range edge-to-edge correlations continue to indicate the presence of SSB in the HO phase for smaller values of $\lambda$. In a) and c) we considered a chain of $L = 51$ spins and $50$ links in between, and set $h_\tau/J_\tau = h_S/J_\tau = 0.1$; gray circles indicate the underlying data points. See Appendix~\ref{sec:appB} for more details on our numerical simulations. 
        }  
	\label{fig:pd_lamda}
\end{figure}

(i) Hidden order HO-SPT phase: $M^*_\tau, M_{S}^*>0$, for $h_\tau/J_\tau<1/2$ and $h_S/J_S<1/2$.
In this phase the link variables $\hat{\tau}^z$ fluctuate strongly, while the spins $\hat{S}^z$ exhibit SSB and the associated long-range order in squeezed space. In the original basis (referred to as real space in the following) spin-correlations are hidden, with $M_S = \langle |\frac{1}{L}\sum_j \hat S^z_j|\rangle=0$ due to the proliferation of domain walls tied to negative links, $\hat{\tau}^z = -1$. Since $[\prod_{j=0}^{L-2} \hat{\tau}^z_{\langle j,j+1 \rangle},\hat{H}]=0$ is conserved, with $L$ the system size, edge-to-edge correlations $\langle \hat{S}^z_0 \hat{S}^z_{L-1} \rangle \equiv \pm  \langle  \hat{S}^z_0 \left( \prod_{j=0}^{L-2} \hat{\tau}^z_{\langle j,j+1 \rangle} \right) \hat{S}^z_{L-1} \rangle$ correspond to non-local string correlations and become long-ranged, as a direct manifestation in real space of the hidden order in squeezed space. When $h_S=0$, the ground state exactly satisfies the \emph{hidden-order rule}, i.e. $\hat{S}^z_j=\left( \prod_{0 \leq i<j} \hat{\tau}^z_{\langle i,i+1 \rangle} \right) \hat{S}^z_0$, and the HIO model exhibits Hilbert space fragmentation~\cite{Rakovszky2020}. Excitations correspond to spin flips not accompanied by changes in the link variables, see Fig.~\ref{fig:1d_pd}~b).

(ii) Link-ordered phase (SRO): $M^*_\tau > 0, M^*_S = 0$, for $h_\tau/J_\tau<1/2$ and $h_S/J_S>1/2$.
This phase has no SSB and lacks long-range order in the spin variables, both in real and squeezed space; the link variables $\hat{\tau}^x$ feature long-range correlations. As in the HO phase, $M_S=0$, and this phase is separated from the HO phase by a hQCP or SPT transition [blue dashed line in Fig.~\ref{fig:1d_pd} c)]: across the hQCP, no local bulk order parameter of the spins $\hat{\vec{S}}_j$ can detect the transition in real space.

(iii) Fully disordered phase (SRO): $M^*_\tau = M^*_S = 0$, for $h_\tau/J_\tau > 1/2$ and $h_S/J_S > 1/2$. 
This completely symmetric phase breaks none of the two $\mathbb{Z}_2$ symmetries ($\hat{\tau}^x / \hat{S}^z \to - \hat{\tau}^x / \hat{S}^z$), implying $M_S=M_\tau=0$. It is separated from the link-ordered phase via a QCP associated with the breaking of the link-$\mathbb{Z}_2$ symmetry [light gray line in Fig.~\ref{fig:1d_pd} c)]. This QCP disappears when the link-$\mathbb{Z}_2$ symmetry is explicitly broken in the Hamiltonian, e.g. by $b_\tau \neq 0$, in which case the link-ordered and fully disordered phases combine into one. 

(iv) Long-range ordered (LRO) phase: $M^*_\tau = 0, M^*_S > 0$, for $h_\tau/J_\tau > 1/2$ and $h_S/J_S < 1/2$. 
This phase spontaneously breaks the spin-$\mathbb{Z}_2$ symmetry, but remains link-$\mathbb{Z}_2$ symmetric. Since flipped $\hat{\tau}^z$ links remain confined~\cite{Endres2011}, the non-zero magnetization in squeezed space, $M^*_S >0$, also manifests in long-range correlations and $M_S>0$ in real space. Transitions into the LRO phase, both from the HO and fully disordered phases, constitute QCPs of Ising type.

The HO phase we find in the 1D HIO model can be understood as an SPT phase, protected by the global $\mathbb{Z}_2$ symmetry $\hat{S}^z_j \to - \hat{S}^z_j$ that is spontaneously broken in squeezed space. This demonstrates that the exact unitary transformation $\hat{U}$, decoupling spins and link variables in squeezed space for $\lambda=0$, is not necessary (though helpful) to observe the HO phase. Indeed, for $\lambda \neq 0$, the unitary $\hat{U}$ does not lead to an exact decoupling, but the HO phase still exists even for values of $\lambda$ close to unity, as we demonstrate in Fig.~\ref{fig:pd_lamda}. 

Similarly, the hQCP between the fully symmetric and the HO phases remains robust when $\lambda \neq 0$. In the 1D HIO model, the hQCP is accompanied by the emergence of link order of the $\hat{\tau}^x$ field, with $M_\tau > 0$ for $h_S/J_S>1/2$. However this is an artifact of the additional global $\mathbb{Z}_2$ symmetry of the link variables. It is absent if the latter is explicitly broken, or, as we will show next, when generalizing the HIO model to 2D.

% % % % % % % % % % % % % % % % % % % % % % % % % % % % 
\section*{Hidden order in 2D: Discrete SSB\\ and finite-$T$ SPT}
\label{sec:2d_HO}
% % % % % % % % % % % % % % % % % % % % % % % % % % % % 
Now we extend the construction of HO and hSSB to higher dimensions. Since SSB at temperatures $T>0$ can only take place in 2D and higher, this opens up the possibility of a finite-temperature hidden critical point (hCP). Indeed, by considering a generalization of the 2D TFIM, we will now explicitly construct a 2D Ising-type hCP, realizing a finite-$T$ SPT transition protected by a 1-form symmetry~\cite{Roberts2017}. The central idea is to decorate domain walls of the Ising spins $\hat{S}^z_{\vec{j}}$ with flipped links $\hat{\tau}^z_{\nn}=-1$, as in 1D. To guarantee that no frustrated links appear, we further ensure that the flipped link variables $\hat{\tau}^z_{\nn}=-1$ form closed loops without ends, realizing the required 1-form symmetry. I.e., the link fields $\hat{\tau}$ need to be described by a loop gas model, the simplest instance of which is Kitaev's toric code~\cite{Trebst2007,KITAEV2003}.

\emph{Squeezed space and HIO model in 2D.--}
This leads us to the 2D HIO Hamiltonian,
\begin{multline}
	\hat H = - J_S \sum_{\nn} \hat{S^z_{\vec{i}}} \hat{S^z_{\vec{j}}} \hat \tau^z_{\nn} + h_S \sum_{\vec{j}} \hat S^x_{\vec{j}}\\
     - h_\tau \sum_l  \hat \tau^z_l
    - \mu_\tau \sum_{\square} \prod_{l \in \square} \hat \tau^z_l
    + J_\tau \sum_{\vec{j}} \hat S^x_{\vec{j}} \prod_{l \in +_{\vec{j}}} \hat \tau^x_l \;.
    \label{eq:TC_meets_Ising}
\end{multline}
The first line describes a TFIM with sign-flipped Ising interactions on bonds where $\hat{\tau}^z_{\nn}=-1$. The second line starts with the string tension $h_\tau$, followed by a plaquette term $\propto \mu_\tau$ defined on plaquettes $\square$ which penalizes open strings. Finally a correlated fluctuation of spins and links $\propto J_\tau$ is added, involving a product over links $l$ forming a star $+_{\vec{j}}$ around site $\vec{j}$. This last term guarantees flipped spins $\hat{S}^z_{\vec{j}}$ to be accompanied by a flip of all links $\hat{\tau}^z_l$ surrounding site $\vec{j}$. The model is illustrated on the square lattice in Fig.~\ref{fig:1d_pd} a).

The 2D HIO model features a global $\mathbb{Z}_2$ symmetry, $\hat{S}^z \to - \hat{S}^z$ that can be spontaneously broken. Restricted to a 1D chain, the 2D HIO model reduces to the 1D HIO model Eq.~\eqref{eq:1d_ham} (for $\lambda=\mu_\tau=0$) if the four-link operators are reduced to two-link terms; this is illustrated in Fig.~\ref{fig:1d_pd} a) and b). For simplicity we only consider the case $\lambda=0$ here, but a generalization of the Ising interactions to $\lambda \in [0,1]$ as in 1D, Eq.~\eqref{eq:1d_ham}, is also possible. Like the 1D HIO model, the 2D HIO model features a second global $\mathbb{Z}_2$ symmetry, $\hat{\tau}^x \to - \hat{\tau}^x$, but this symmetry cannot be broken spontaneously because it turns into a local $\mathbb{Z}_2$ gauge symmetry of the loop gas model in Eq.~\eqref{eq:TC_meets_Ising} that will be discussed further below. 

To derive the phase diagram and solve the 2D HIO model, Eq.~\eqref{eq:TC_meets_Ising}, we apply a similar strategy as in 1D and construct a unitary $\hat{U}$ that disentangles spins and link variables. This only works, however, when the strings defined by $\hat{\tau}^z_l=-1$ form \emph{closed} loops, i.e. for
\begin{equation}
    \prod_{l \in \square} \hat \tau^z_l \ket{\psi} \equiv \hat{B}_\square \ket{\psi}= \ket{\psi}, \quad \forall ~\square.
    \label{eqClosedLoopSubspace}
\end{equation}
Since $[\hat{B}_\square,\hat{H}]=0$ this defines a sector of the HIO Hilbert space, realized by $\mu_\tau \to \infty$, to which we will restrict ourselves in the following. In the $T=0$ ground state it is sufficient to assume $\mu_\tau > 0$, which leads to a ground state of Eq.~\eqref{eq:TC_meets_Ising} in the closed-loop sector. 

The unitary transformation $\hat{U}$ defining squeezed space in 2D, is the same as in 1D, see Eq.~\eqref{eq:1d_HO_unitary}. It is only in the definition of $\hat p_{\vec{j}}$ that care has to be taken: We define 
\begin{equation}
    (-1)^{\hat{p}_{\vec{j}}}=\prod_{l\in \mathcal{L}_{\vec{j}}} \hat \tau_l^z
    \label{eqDefString}
\end{equation}
as a product of link variables along a path $\mathcal{L}_{\vec{j}}$ from some fixed reference site to $\vec{j}$. For the different sites $\vec{j}$ we choose a path $\mathcal{L}_{\vec{j}}$ following a one-dimensional snake-like covering of all lattice sites, as described in Appendix~\ref{sec:appA}. Since this parity $\hat{p}_{\vec{j}}$ is independent of the path $\mathcal{L}_{\vec{j}}$ in the subspace of closed loops, Eq.~\eqref{eqClosedLoopSubspace}, the above expression for $(-1)^{\hat{p}_{\vec{j}}}$ is well-defined. Intuitively, the parity $\pi_{\vec{j}}=(-1)^{p_{\vec{j}}}$ distinguishes sites $\vec{j}$ inside ($\pi_{\vec{j}}=-1$) and outside ($\pi_{\vec{j}}=+1$) of the closed loops of strings $\tau^z=-1$. By applying the unitary $\hat{U}$ in Eq.~\eqref{eq:1d_HO_unitary}, spins inside closed loops are flipped: This is the defining property of squeezed space in 2D.

Now we apply $\hat{U}$ to the 2D HIO Hamiltonian in the closed-loop subspace, which decouples the system into a TFIM of spins $\hat{\vec{S}}$ and a toric code in a field (TC-F), see Appendix~\ref{sec:appA} for a derivation,
\begin{multline}
	\hat{U}^\dagger \hat{H} \hat{U} =  \underbrace{- J_S \sum_{\nn} \hat{S}^z_{\vec{i}} \hat{S}^z_{\vec{j}} + h_S \sum_{\vec{j}} \hat{S}^x_{\vec{j}}}_{\hat{H}_{\rm TFIM}}\\
    + \underbrace{\frac{J_\tau}{2} \sum_{\vec{j}} \prod_{l \in +_{\vec{j}}} \hat \tau^x_l
     - h_\tau \sum_l  \hat \tau^z_l}_{\hat H_{\rm TC-F}}\;.
    \label{eq:decoupling}
\end{multline}
Note that we dropped the term $\propto \mu_\tau$ from Eq.~\eqref{eq:TC_meets_Ising}, since we work in the sector $\hat{B}_\square=1$ where it becomes a constant. Similar transformations have recently been discussed by Ref.~\cite{Rakovszky2023}.

\emph{Zero-temperature phase diagram.--}
As in the 1D HIO model, the decoupling of the spin and link degrees of freedom results in a factorization of eigenstates in squeezed space. The zero-temperature phase diagram is similar to the 1D case, shown in  Fig.~\ref{fig:1d_pd} c), with independent, straight phase boundaries. For $h_S/J_S < (h_S/J_S)_{\rm c,2D}$ the global $\mathbb{Z}_2$ symmetry is spontaneously broken, $M_S^*>0$, manifesting in hidden and long-range order, respectively, depending on the loop gas configuration. For larger values of $h_S/J_S$, the spins realize a $\mathbb{Z}_2$ symmetric paramagnet with $M_S=M_S^*=0$.

The most interesting phase is the HO phase, in which long-range order in squeezed space, $M_S^* >0$, is hidden in real space by fluctuating strings, $M_S=0$. This happens for $h_\tau/J_\tau<(h_\tau/J_\tau)_{\rm c, 2D}$ when the loop gas is deconfined (topologically non-trivial) and strings $\hat{\tau}^z=-1$ percolate through the entire system~\cite{Linsel2024PRB, Duennweber2025, Linsel2025arXiv}. The most direct way to understand this SPT phase comes from the limit $h_S=0$: In this case, spins $\hat{S}^z$ are fully polarized in squeezed space; in real space, the hidden-order rule relates spins to links through 
\begin{equation}
    \hat{S}^z_{\vec{j}} = \hat{S}^z_{\vec{r}} \prod_{l\in \mathcal{L}_{\vec{j}}} \hat \tau_l^z, \qquad \text{for}~h_S=0,
    \label{eqHOrule}
\end{equation}
where $\vec{r}$ is the reference site to which $\mathcal{L}_{\vec{j}}$ connects site $\vec{j}$. I.e., the spins $\hat{S}^z$ realize the dual variables of the loop gas~\cite{Peierls1936,Wegner1971}. The latter undergo an Ising transition as $h_\tau/J_\tau$ is increased beyond $(h_\tau/J_\tau)_{\rm c, 2D}$, restoring long-range order in real space, $M_S>0$, in the confined (topologically trivial) phase of the loop gas where strings form finite-size loops and do not percolate.

The transition from the HO phase to the disordered phase realizes a hQCP, or a 2D SPT transition. It is invisible to local order parameters in real space, since $M_S=0$ remains zero; i.e., both sides of the transition appear symmetric in their bulks. The hQCP can be detected directly in squeezed space or, equivalently, via a string order parameter 
\begin{equation}
    C^*(\vec{j}) = \bigg\langle \hat{S}^z_{\vec{r}} \bigg( \prod_{l\in \mathcal{L}_{\vec{j}}} \hat \tau_l^z \bigg) \hat{S}^z_{\vec{j}} \bigg\rangle.
    \label{eqCstring2D}
\end{equation}
Along edges of the system without open strings (dangling bonds) and still considering the closed-loop subspace $\mu_\tau \to \infty$, the correlator $C^*(\vec{j})$ develops long-range edge correlations $C_{\rm e}(d_{\rm e}) = \langle \hat{S}^z_{\vec{r}_{\rm e}} \hat{S}^z_{\vec{r}_{\rm e}+\vec{d}_{\rm e}} \rangle$ in the HO phase, demonstrating the SPT nature of the latter. These long-range edge correlations require the 1-form symmetry underlying the closed-loop condition to remain stable. 

Finally, within the $\mathbb{Z}_2$ symmetric, disordered phase for $h_S/J_S > (h_S/J_S)_{\rm c,2D}$, another topological phase transition takes place at $(h_\tau/J_\tau)_{\rm c, 2D}$: This is the confining transition of the loop gas, which can be detected through a Wilson loop~\cite{Fradkin1979} or a percolation analysis of link snapshots~\cite{Linsel2024PRB}, but has no influence on the spins $\hat{\vec{S}}$.

\emph{Finite-temperature phase diagram.--}
Next, we turn to the finite-temperature phase diagram of the 2D HIO model, which is a lot more interesting than in 1D since the discrete $\mathbb{Z}_2$ symmetry $\hat{S}^z \rightarrow - \hat{S}^z$ can be broken at $T>0$. In the following we assume $\mu_\tau \to \infty$, restricting our entire model to the closed loop subspace\footnote{Note that in an experimentally relevant, finite-size setup it is sufficient to choose $\mu_\tau$ large enough to suppress the total number of open-loop defects to below one on average. This requires $\mu_\tau \to \infty$ as the linear system size $L\to \infty$.}, Eq.~\eqref{eqClosedLoopSubspace}. Hence, the decoupled Hamiltonian in squeezed space, Eq.~\eqref{eq:decoupling}, remains valid, allowing us to derive the phases of spin and link variables independently. We start with the spins in the 2D TFIM, which spontaneously break the discrete $\mathbb{Z}_2$ symmetry, and form an ordered state in 2D squeezed space signified by $M_S^*(T)>0$, below a critical temperature $T<T_c(J_S,h_S)$ depending on $J_S$ and $h_S$. 

\begin{figure}
	\includegraphics[width=.5\textwidth]{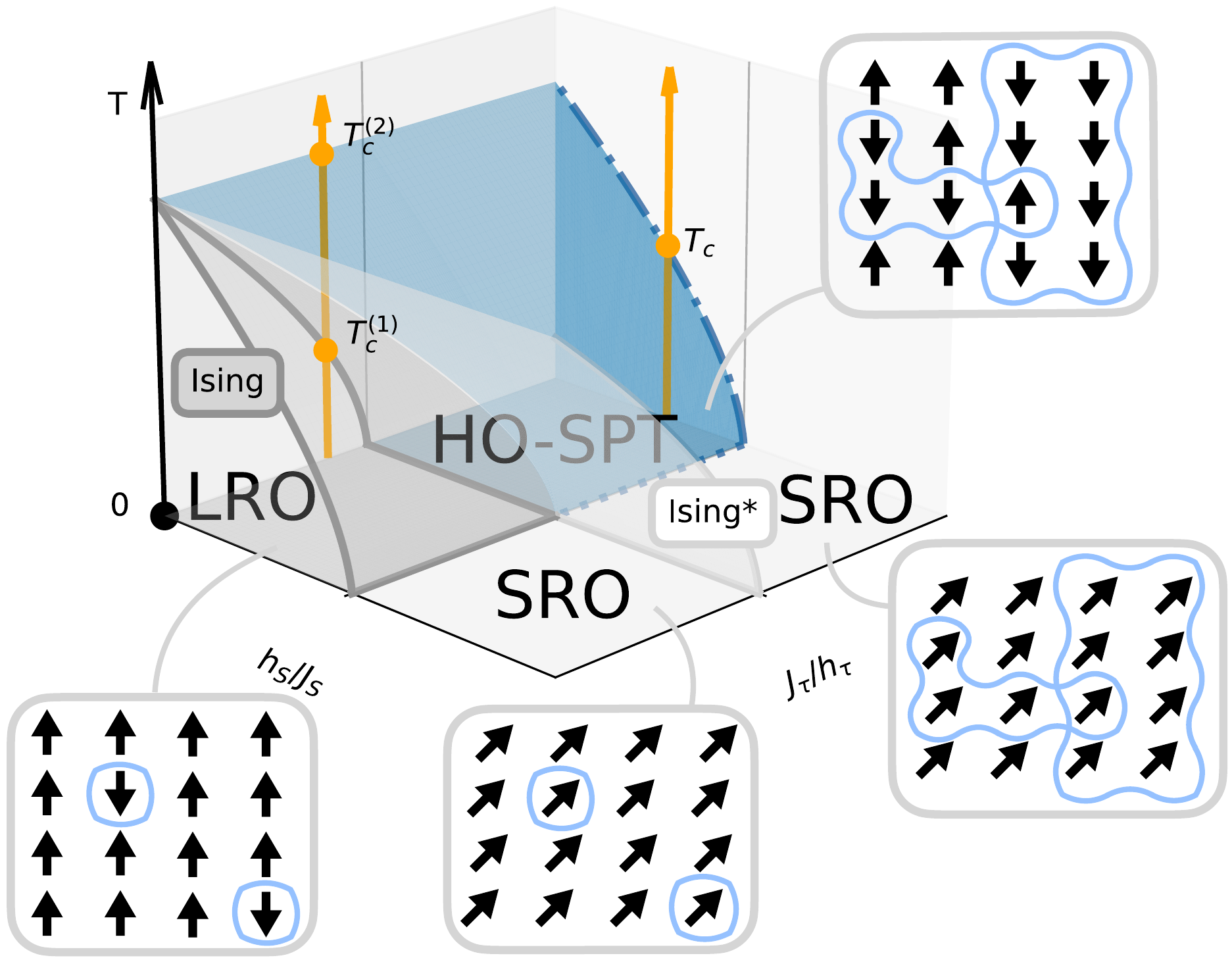}
	\caption{\justifying 
    Schematic phase diagram of the 2D HIO model, Eq.~\eqref{eq:TC_meets_Ising}, at finite temperature and in the closed-loop limit, $\mu_\tau \to \infty$. The decoupling of spin and link degrees of freedom in squeezed space, after applying the unitary transformation in Eq.~\eqref{eq:1d_HO_unitary}, leads to a factorization of the phase diagram. The insets illustrate the respective system configurations. The TFIM of spins $\hat{\vec{S}}$ features a finite-$T$ Ising transition, describing where SSB takes place. The toric code in a field describing links $\hat{\vec{\tau}}$ is dual to a TFIM, with an associated ${\rm Ising}^*$ transition characterizing the deconfinement of the loop gas. In regimes where the loop gas is deconfined (percolating), through quantum or thermal fluctuations, spin order is hidden (blue region) in real space. For large $J_\tau / h_\tau$, a finite-$T$ SPT transition at $T_c$ is obtained. In the confined (non-percolating) region of the loop gas, for small $J_\tau / h_\tau$, and when $h_S/J_S$ is small, low-$T$ LRO gives way to a HO phase at $T_c^{(1)}$ where the loop gas thermally deconfines, before entering the fully symmetric, disordered phase at $T_c^{(2)}$. In this regime, the critical temperature separating the ordered and disordered phase is $T/J_S \approx 2.27$ at elevated temperatures.
    }
	\label{fig:2d_T}
\end{figure}

The phase diagram of the closed-loop gas model at $T>0$ in squeezed space is similar. Through the help of a duality mapping, the toric code in a field is equivalent to a 2D TFIM, with Ising interactions $h_\tau$ and a transverse field $J_\tau$~\cite{Wegner1971}. Below $T_c(J_\tau,h_\tau)$, in the ordered phase of the dual variables, the loop gas forms non-percolating, finite-size clusters of strings. In contrast, above $T_c(h_\tau,J_\tau)$, in the disordered phase of the dual variables, the loop gas forms a percolating net of strings extending across the entire system~\cite{Linsel2024PRB}. The respective configurations are indicated in the insets of Fig.~\ref{fig:2d_T}. Notably, SSB of the dual variables in their ordered phase has no equivalent in the original string basis -- in contrast to SSB of the spins $\hat{\vec{S}}$, which leads to e.g. a doubly-degenerate ground state. For both models in squeezed space, the respective $T_c=0$ vanishes at the zero-temperature quantum phase transitions at $(J_S/h_S)_{\rm c, 2D}$ and $(J_\tau/h_\tau)_{\rm c, 2D}$.

The resulting finite-temperature phase diagram of the closed-loop 2D HIO model is shown in Fig.~\ref{fig:2d_T}. One of its most interesting features is the extension of the HO, SPT-type phase above $T>0$. For small $h_S/J_S$, ensuring $M_S^*>0$, and provided the loop gas is in its percolating phase, i.e. for large $J_\tau/h_\tau$ and ensuring that $M_S=0$, we find that the $T=0$ HO phase extends to some $T_c>0$. At this $T_c$, it turns into the fully symmetric, disordered phase, realizing a \emph{finite-$T$ SPT transition}.

For smaller values of $J_\tau/h_\tau$, where the loop gas is in its non-percolating, confined phase at low $T$, the spontaneous breaking of the spin's $\mathbb{Z}_2$ symmetry manifests in long-range order, $M_S>0$. This is stable up to a critical temperature $T_c^{(1)}$, above which $M_S=0$. When simultaneously $h_S/J_S$ is sufficiently small, we obtain a second critical $T_c^{(2)}$: In between, for $T_c^{(1)} < T < T_c^{(2)}$, the $\mathbb{Z}_2$ symmetry remains broken but the system is in the HO-SPT phase. Here, thermal fluctuations induce fluctuations of the link variables which gives rise to a thermally restored SPT phase~\cite{Tiwari2024}. Only beyond $T>T_c^{(2)}$ the $\mathbb{Z}_2$ symmetry is restored. Thereby we establish an interesting new scenario how long-range order can be destroyed in a step-like manner as temperature is increased, from LRO to HO and finally to the disordered phase in an SPT transition. The two scenarios are illustrated in Fig.~\ref{fig:2d_T} along exemplary scans through the phase diagram (orange arrows).

\emph{Open strings \& relation to Ising gauge theory.--}
So far we restricted our discussion of the 2D HIO model to the subspace of closed $\hat{\tau}^z$ loops. Next, we consider finite $\mu_\tau < \infty$ and include an additional term in the Hamiltonian introducing $\hat{\tau}^z$ strings with open ends:
\begin{equation}
    \hat{H} \rightarrow \hat{H} + h_X \sum_l \hat{\tau}^x_l.
    \label{eqAddhX}
\end{equation}
With this term included, we can no longer use the unitary $\hat{U}$ to decouple spin and link variables, since the inside and outside of the $\hat{\tau}^z$ loops become ill-defined in the presence of open strings.

\begin{figure}
	\includegraphics[width=.42\textwidth]{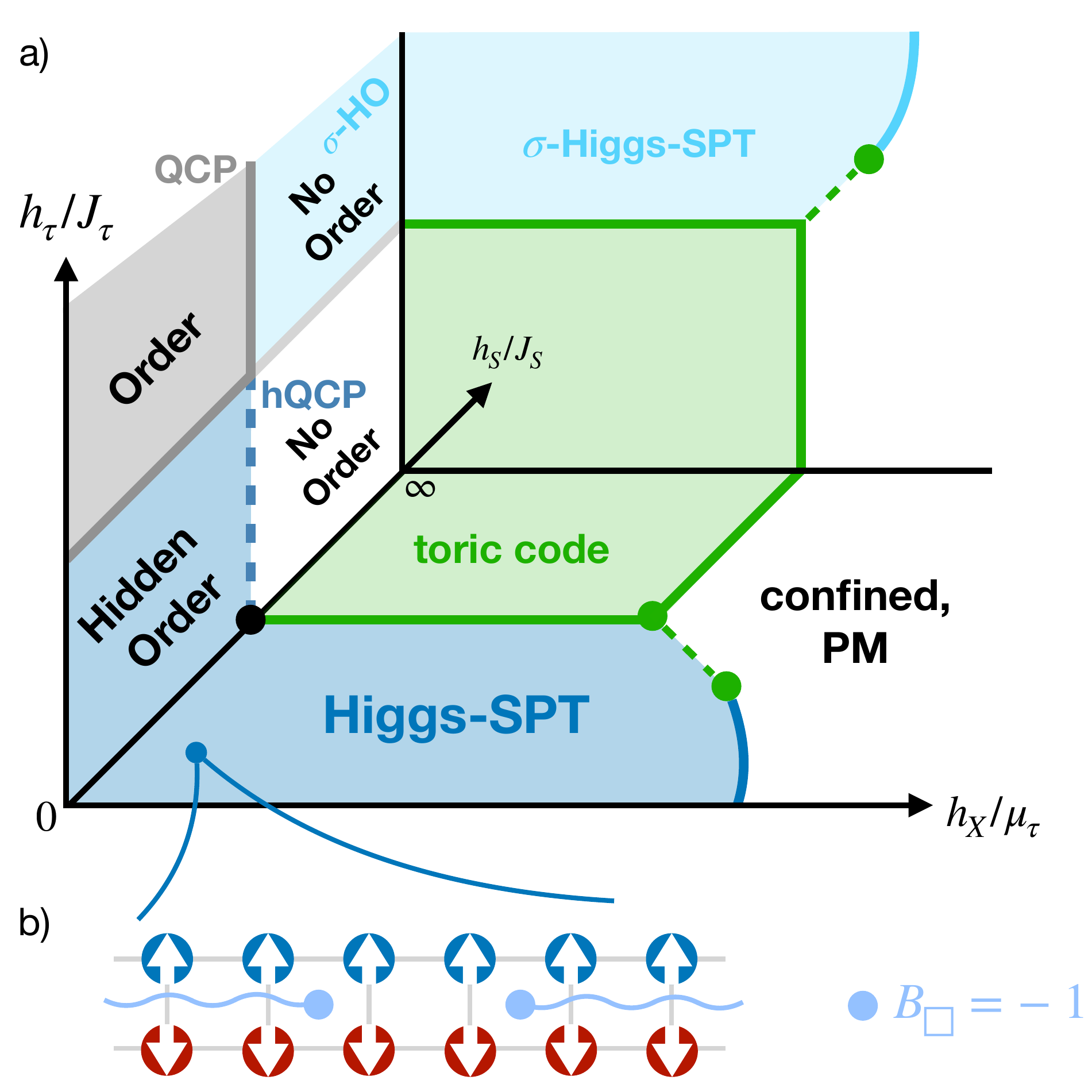}
	\caption{\justifying 
    Zero-temperature phase diagram of the double-Higgs $\mathbb{Z}_2$ gauge theory, Eq.~\eqref{eqIGT}. a) On the left vertical plane, for $h_X=0$, the exactly solvable phases of the 2D HIO model are obtained, including HO of both Higgs fields. The latter regimes connect directly to the respective Higgs-SPT phases of $S$ ($\sigma$) on the bottom horizontal (back vertical) plane. The phase boundaries of these SPT phases - which we conjecture to coincide with our HO-SPT phases - are taken from Ref.~\cite{Verresen2024}. The topological toric code phase (green) corresponds to the disordered, PM phases of the Higgs fields. b) The HO phase remains stable upon including open $\tau^z$ strings, by $h_X \neq 0$, because domain walls of $\hat{S}^z$ spins lead to a linear confining force between open ends of the strings, $B_\square = -1$.}
	\label{fig:2d_ho_tau}
\end{figure}

We will argue next that the HO phase remains stable even when $h_X\neq 0$ and open $\hat{\tau}^z$ strings are included. By relating the HIO model to a double-Higgs Ising gauge theory (IGT)~\cite{Verresen2024,Rakovszky2023}, we will show that the HO phase coincides with the Higgs phase of the IGT. Indeed, it was recently shown at zero temperature that the Higgs phase realizes an SPT phase~\cite{Verresen2024}: We conjecture that the HO-SPT phase we constructed above is identical to the Higgs-SPT phase found by Verresen et al., see Fig.~\ref{fig:2d_ho_tau} a), and extends to $T>0$.

Before going into details, we provide intuition why the HO phase remains stable upon including open strings, for $h_X>0$. To this end, we consider the limit $h_S=0$, where the hidden-order rule, Eq.~\eqref{eqHOrule}, applies when $h_X=0$. I.e., any string segment $\tau^z_{\nn}=-1$ is bound to a domain wall of the Ising spins, $S^z_{\vec{i}}=-S^z_{\vec{j}}$. Adding small $|h_X| \ll \mu_\tau$ can perturbatively open the string, but keeps the spin-domain wall unchanged, as we illustrate in Fig.~\ref{fig:2d_ho_tau} b). This costs energy $\propto J_s$ per open string segment, and realizes a force linear in the distance between the two charges, $B_\square=-1$, at the open ends of the strings. As long as these charges remain confined, the inside and outside of the loop gas can still be meaningfully defined and HO is stabilized. When $h_X/\mu_\tau$ becomes too large, open ends with $B_\square=-1$ proliferate and deconfine, destroying the HO phase in an SPT transition at $h_{X,c}>0$. 

Now we proceed by describing the 2D HIO model in the framework of an IGT. When the string tension associated with the links vanishes, $h_\tau=0$, the Hamiltonian Eq.~\eqref{eq:TC_meets_Ising} features a further local Gauss law, $[\hat H, \hat{G}_{\vec{j}}] = 0$ with $\hat G_{\vec{j}} = \prod_{l \in +_{\vec{j}}} \hat\tau^x_l \hat{S}^x_{\vec{j}}$, in addition to the closed-loop constraint, Eq.~\eqref{eqClosedLoopSubspace}. By introducing a second Higgs field $\hat{\sigma}^x$, in addition to $\hat{S}^z$, to describe the open ends of the $\hat \tau^x$ strings, we can elevate the entire HIO model, for any $h_\tau$, to an IGT:
\begin{multline}
	\hat H_{\rm IGT} = - J_S \sum_{\nn} \hat{S}^z_{\vec{i}} \hat{S}^z_{\vec{j}} \hat \tau^z_{\nn} + h_S \sum_{\vec{j}} \hat{S}^x_{\vec{j}} + h_X \sum_l \hat{\tau}^x_l\\
     - h_\tau \sum_{\nn} \hat{\sigma}^z_{\vec{i}} \hat{\sigma}^z_{\vec{j}} \hat{\tau}^z_{\nn} 
     + J_\tau \sum_{\vec{j}} \hat{\sigma}^x_{\vec{j}} 
    - \mu_\tau \sum_{\square} \prod_{l \in \square} \hat{\tau}^z_l\;.
    \label{eqIGT}
\end{multline}
This Hamiltonian acts in a Hilbert space satisfying the following Gauss law,
\begin{equation}
    2 \hat{S}^x_{\vec{j}} \hat{\sigma}^x_{\vec{j}} \prod_{l \in +_{\vec{j}}} \hat{\tau}^x_l \ket{\psi} \equiv \hat{G}_{\vec{j}} \ket{\psi} = \ket{\psi}, \quad \forall ~ \vec{j}.
    \label{eqIGTGauss}
\end{equation}

The $T=0$ phase diagram of the double-Higgs IGT, Eq.~\eqref{eqIGT}, is shown in Fig.~\ref{fig:2d_ho_tau} a). We construct it starting from the phase diagram of the 2D HIO, see Fig.~\ref{eq:TC_meets_Ising} a), at $h_X/\mu_\tau=0$. Next, we set $h_\tau=0$ ($J_\tau\gg0$) in Eq.~\eqref{eqIGT} and eliminate $\hat{S}^x$ making use of the Gauss law, Eq.~\eqref{eqIGTGauss}: $\hat{S}^x_{\vec{j}} = \frac{1}{2}\hat{\sigma}^x_{\vec{j}} \prod_{l \in +_{\vec{j}}} \hat{\tau}^x_l$. The resulting Hamiltonian commutes with $\hat{\sigma}^x_{\vec{j}}$, which takes the value $\sigma^x_{\vec{j}}=-1$ in the ground state of the link variables. This finally leads to the identification $\hat{S}^x_{\vec{j}} = - \frac{1}{2} \prod_{l \in +_{\vec{j}}} \hat{\tau}^x_l$ and
\begin{multline}
    \hat{H}(\nicefrac{h_\tau}{J_\tau}=0) = - J_S \sum_{l} \hat \tau^z_l - \frac{h_S}{2} \sum_{\vec{j}} \prod_{l \in +_{\vec{j}}} \hat{\tau}^x_l \\
    + h_X \sum_l \hat{\tau}^x_l - \mu_\tau \sum_{\square} \prod_{l \in \square} \hat{\tau}^z_l,
    \label{eqHIGThtauZro}
\end{multline}
which is the well-known perturbed toric code Hamiltonian with two fields, $J_S$ and $h_X$~\cite{Fradkin1979,Trebst2007}. Its phase diagram is sketched in the $h_\tau=0$ plane in Fig.~\ref{fig:2d_ho_tau} a), demonstrating that the HO phase is directly connected to the Higgs-SPT phase. 

Further insights into the phase diagram of the double-Higgs IGT can be obtained by using the symmetry between the two Higgs fields. Exchanging $2\hat{\vec{S}} \leftrightarrow \hat{\sigma}$, as well as $J_S \leftrightarrow h_\tau$ and $h_S \leftrightarrow J_\tau$, the Hamiltonian $\hat{H}_{\rm IGT}$ is invariant. This establishes the relation between the $\sigma$-HO phase and the $\sigma$-Higgs-SPT phases on the vertical planes in Fig.~\ref{fig:2d_ho_tau} a). The ordered phase at large $h_\tau/J_\tau$ and small $h_S/J_S$ features LRO of both Higgs fields, $\sigma^z$ and $S^z$. Finally, for large $h_X/\mu_\tau$ both Higgs fields are in a fully symmetric, paramagnetic (PM) phase.

In combination with our earlier results, we conclude that the HO / Higgs-SPT phase protected by the global $\mathbb{Z}_2$ symmetry of spins, $\hat{S}^z \rightarrow - \hat{S}^z$, remains stable at finite temperatures, $T>0$. We conjecture that it features hSSB of the global $\mathbb{Z}_2$ symmetry, and thus constitutes an extension of the SSB $\hat{S}^z$-ordered phase to a regime with hidden order. This stability to thermal fluctuations is in stark contrast to the topological toric code phase, which is \emph{not} robust at any $T>0$ due to the emergence of a non-zero density of thermal excitations.

% % % % % % % % % % % % % % % % % % % % % % % % % % % % 
\section*{Hidden order in 2D: Continuous SSB and hidden BKT}
\label{sec:2d_U1_HO}
% % % % % % % % % % % % % % % % % % % % % % % % % % % % 
Next we extend our construction of phases with hidden order to systems with continuous symmetries. To this end we replace the TFIM of the spins $\hat{\vec{S}}$ by the XY or XXZ model featuring a continuous $U(1)$ symmetry. At $T=0$ the latter can be spontaneously broken, and in this regime we will construct a HO / SPT phase featuring a (hidden) gapless Goldstone mode associated with the broken continuous symmetry. This constitutes an intrinsically gapless SPT phase~\cite{Thorngren2020} in two dimensions. At $T>0$, by the Mermin-Wagner-Hohenberg theorem, the $U(1)$ symmetry cannot be spontaneously broken. Instead, the XY model features a topological BKT transition from which we will construct a hidden BKT (or BKT-class SPT) transition at finite $T$. The latter can be characterized by edge correlations, protected by a 1-form symmetry realized through the closed-loop condition, that turn from quasi-long ranged power-law to short-ranged exponential.

\emph{Squeezed space and HXYO model.--}
Our starting point is the following Hamiltonian, which we will refer to as the hidden-XY order (HXYO) model,
\begin{multline}
	\hat H = - \frac{J_S}{2} \sum_{\nn} \left( \hat{S}^+_{\vec{i}} \hat{S}^-_{\vec{j}} \hat{\tau}^z_{\nn} + \hc \right) + \Delta \sum_{\nn} \hat{S}^z_{\vec{i}} \hat{S}^z_{\vec{j}} \\
     - h_\tau \sum_l  \hat \tau^z_l
    - \mu_\tau \sum_{\square} \prod_{l \in \square} \hat \tau^z_l
    + J_\tau \sum_{\vec{j}} \hat S^z_{\vec{j}} \prod_{l \in +_{\vec{j}}} \hat \tau^x_l \;.
    \label{eq:U1_meets_tc}
\end{multline}
It describes spin-1/2 $\hat{\vec{S}}$ with XY interactions $\propto J_S$ coupled to a loop gas model, where links $\tau^z_{
\nn}=-1$ introduce flipped signs. Here $\hat S_{\vec{j}}^\pm = \hat S_\vec{j}^x \pm i \hat S_\vec{j}^y$ denotes the spin raising and lowering operators on site $\vec{j}$. We added Ising interactions $\propto \Delta$ among the spins, which are not affected by the link variables. The second line describes a perturbed toric code, where the last term is a correlated fluctuation of strings $\tau^z=-1$ and spins; note that $\hat{S}^z_{\vec{j}}$ in the last term flips the sign of both $\hat{S}^\pm_{\vec{j}}$.

The HXYO model has a global continuous $U(1)$ symmetry of the spins, $\hat{S}^\pm \to e^{\pm i \varphi} \hat{S}^\pm$, corresponding to rotations around $\hat{S}^z$. This symmetry will protect the hidden order that we describe next. Moreover, it implies that $[\hat{H},\hat{S}^z_{\rm tot}]=0$, i.e. $\hat{H}$ can be solved for every value of $S^z_{\rm tot}=\sum_{\vec{j}
} S^z_{\vec{j}}$ separately. Alternatively, a chemical potential term $\mu_S \sum_{\vec{j}} \hat{S}^z_{\vec{j}}$ can be added to the Hamiltonian and $\mu_{S}$ can be tuned instead of $S^z_{\rm tot}$.

As before, in order to solve the HXYO model we work in the closed-loop subspace obtained in the limit $\mu_\tau \to \infty$. We apply the same unitary transformation $\hat{U}$ as in the HIO model, Eqs.~\eqref{eq:1d_HO_unitary}, \eqref{eqDefString} with $\hat{S}^x$ replaced by $\hat{S}^z$, see Appendix~\ref{sec:appA}, which decouples the HXYO Hamiltonian:
\begin{multline}
    \hat{U}^\dagger \hat{H} \hat{U} = \underbrace{- \frac{J_S}{2} \sum_{\nn} \left( \hat{S}^+_{\vec{i}} \hat{S}^-_{\vec{j}} + \hc \right) + \Delta \sum_{\nn} \hat{S}^z_{\vec{i}} \hat{S}^z_{\vec{j}}}_{= \hat{H}_{\rm XXZ}} \\
     \underbrace{- h_\tau \sum_l  \hat \tau^z_l
    + \frac{J_\tau}{2} \sum_{\vec{j}} \prod_{l \in +_{\vec{j}}} \hat \tau^x_l}_{=\hat{H}_{\rm TC-F}} \;.
    \label{eqUXXZtc}
\end{multline}
The toric code model in the second line is defined in the sector with $B_{\square}=1$.

\begin{figure}
	\includegraphics[width=.5\textwidth]{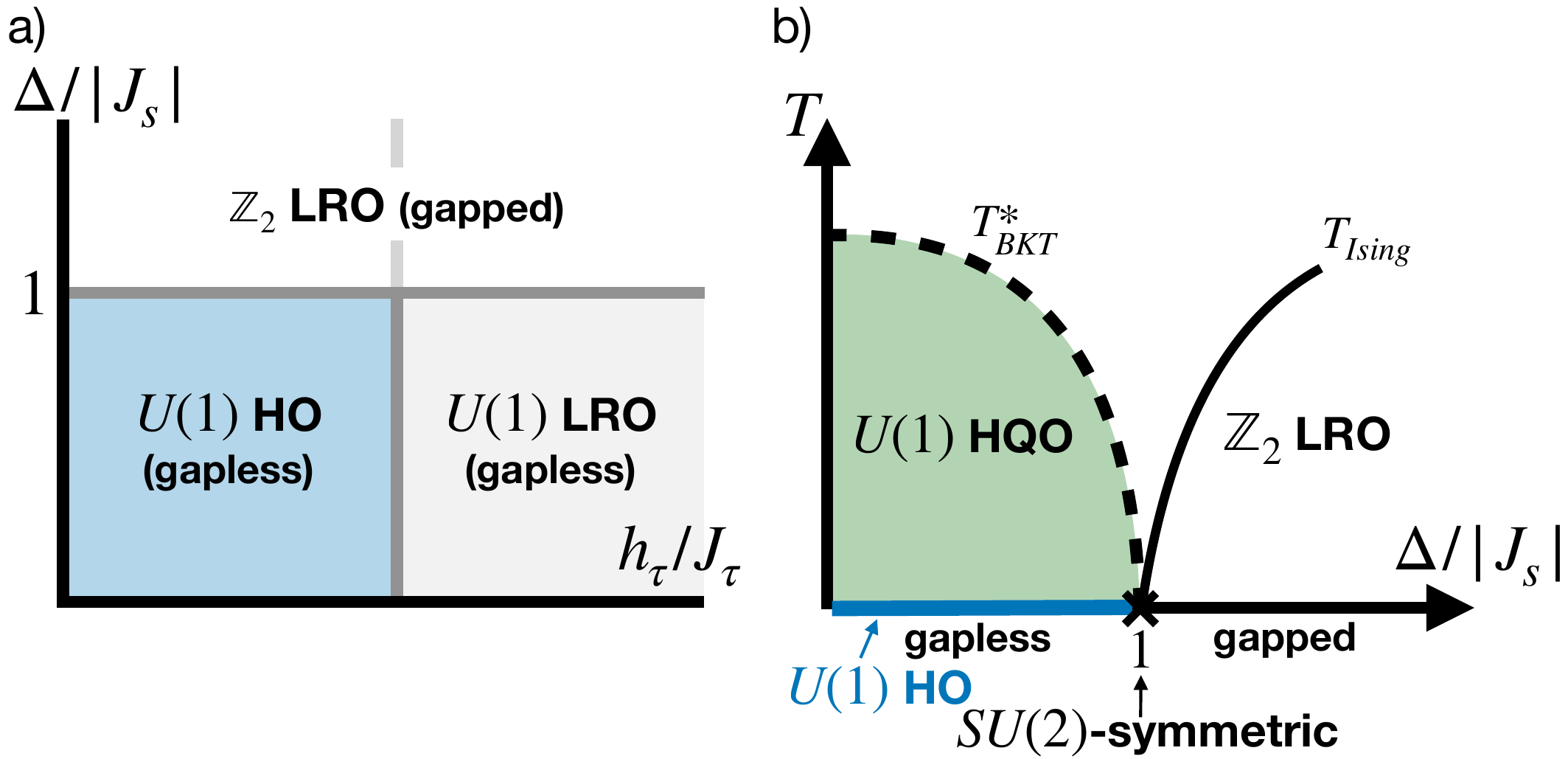}
	\caption{\justifying 
    Phase diagram of the HXYO model, Eq.~\eqref{eq:U1_meets_tc}, in the closed-loop limit $\mu_\tau \to \infty$. a) At zero temperature, long-range $U(1)$ (or XY) order associated with a spontaneously broken $U(1)$ symmetry develops in squeezed space when $|\Delta|<|J_S|$. This manifests in a gapless SPT phase with $U(1)$ hidden order (HO) for small $h_\tau/J_\tau$, and turns into a conventional $U(1)$-ordered phase when $\tau^z$ strings confine at large $h_\tau$. For $|\Delta|>|J_S|$, a $\mathbb{Z}_2$ ordered phase in the original basis is obtained, independent of the loop gas. b) At finite temperature, the hidden $U(1)$ order turns into hidden quasi-long range order, with power-law correlations in squeezed space. At higher temperatures $T^*_{\rm BKT}$ a hCP / finite-$T$ SPT transition into a symmetric, paramagnetic phase is found. The $\mathbb{Z}_2$ order remains stable up to $T_{\rm Ising}$ where it disappears in a finite-$T$ symmetry-breaking transition of Ginzburg-Landau type, in the Ising universality class. At $|\Delta|=|J_S|$, a hidden $SU(2)$ symmetry precludes any finite-$T$ phase transitions.
    }
	\label{fig:2dHXYO}
\end{figure}

\emph{Zero-temperature phase diagram.--}
From the exact decoupling of spins and links in Eq.~\eqref{eqUXXZtc} we obtain again a typical hidden-order phase diagram as in Fig.~\ref{fig:1d_pd} c), with two orthogonal phase boundaries, see Fig.~\ref{fig:2dHXYO} a). For small $|\Delta| < |J_S|$, an XY-phase with a spontaneously broken continuous $U(1)$ symmetry is formed in squeezed space. When $h_\tau < J_\tau$ is small, strings formed by links $l$ with $\tau^z_l=-1$ percolate through the system, suppressing the long-range XY correlations in squeezed space in the original basis: this leads to a hidden-XY, or $U(1)$, ordered phase. It can be characterized by a non-zero string order parameter of the form $C^*(\vec{j}) = \langle \hat{S}^+_{\vec{r}} (\prod_{l \in \mathcal{L}_{\vec{j}}} \hat{\tau}^z_l) \hat{S}^-_{\vec{j}} \rangle$, similar to the one in Eq.~\eqref{eqCstring2D}.

When $h_\tau/J_\tau$ increases and reaches a critical value $(h_\tau/J_\tau)_{\rm c,2D}$, the $\tau^z$ strings confine and stop to percolate through the entire system. In this regime, the long-range XY order associated with the spontaneously broken $U(1)$ symmetry also manifests as long-range order, $\langle \hat{S}^+_{\vec{i}} \hat{S}^-_{\vec{j}} \rangle \to {\rm const.}$ as $|\vec{i}-\vec{j}| \to \infty$, in the original basis. This corresponds to the $U(1)$-ordered phase in Fig.~\ref{fig:2dHXYO} a).

For $|\Delta| > |J_S|$, the spins in the XXZ model transition to a gapped, $U(1)$ symmetric phase. For $\mu_S=0$ this phase is not entirely trivial, because it spontaneously breaks an additional, discrete $\mathbb{Z}_2$ symmetry associated with the reversal of $S^z$ spins, leading to an Ising (anti-) ferromagnet for $\Delta < 0 (>0)$. Since $S^z$ interactions $\propto \Delta$ decouple from the link degrees of freedom in the HXYO model, the $\mathbb{Z}_2$ order associated with this phase can be detected in real and squeezed space alike. Moreover, it is unaffected by the confinement transition of the loop gas model which only depends on the ratio $h_\tau/J_\tau$.

Next, we discuss the excitation spectrum of the HXYO model. Because excitation energies are invariant under the unitary $\hat{U}$, we immediately conclude that the loop gas sector is always gapped, except at the transition point $(h_\tau/J_\tau)_{\rm c,2D}$. In contrast, the XY phase of the XXZ model features a gapless Goldstone mode, for $|\Delta| < |J_S|$, demonstrating that the hidden $U(1)$ ordered phase represents a class of intrinsically gapless SPT states. 

Understanding the spectral weight of the low-energy Goldstone boson requires more care, because the unitary transformation $\hat{U}$ entangles spin and link variables. Since the Goldstone mode can be viewed as a combination of $S^\pm$ operators, $\hat{\gamma} \sim u \hat{S}^- + v \hat{S}^+$, it turns into a non-local string operator in the original basis, 
\begin{equation}
    \hat{U} ~ \hat{\gamma} ~ \hat{U}^\dagger \sim \left( \prod_{l \in \mathcal{L}} \hat{\tau}^z_l \right) \hat{\gamma}.
\end{equation}
Thus, in general, local operators in the original basis couple to collective excitations of both the link and the spin sectors, complicating the direct detection of the gapless SPT Goldstone boson we predict. A detailed discussion will be devoted to future work. 

\begin{figure*}
	\includegraphics[width=.85\textwidth]{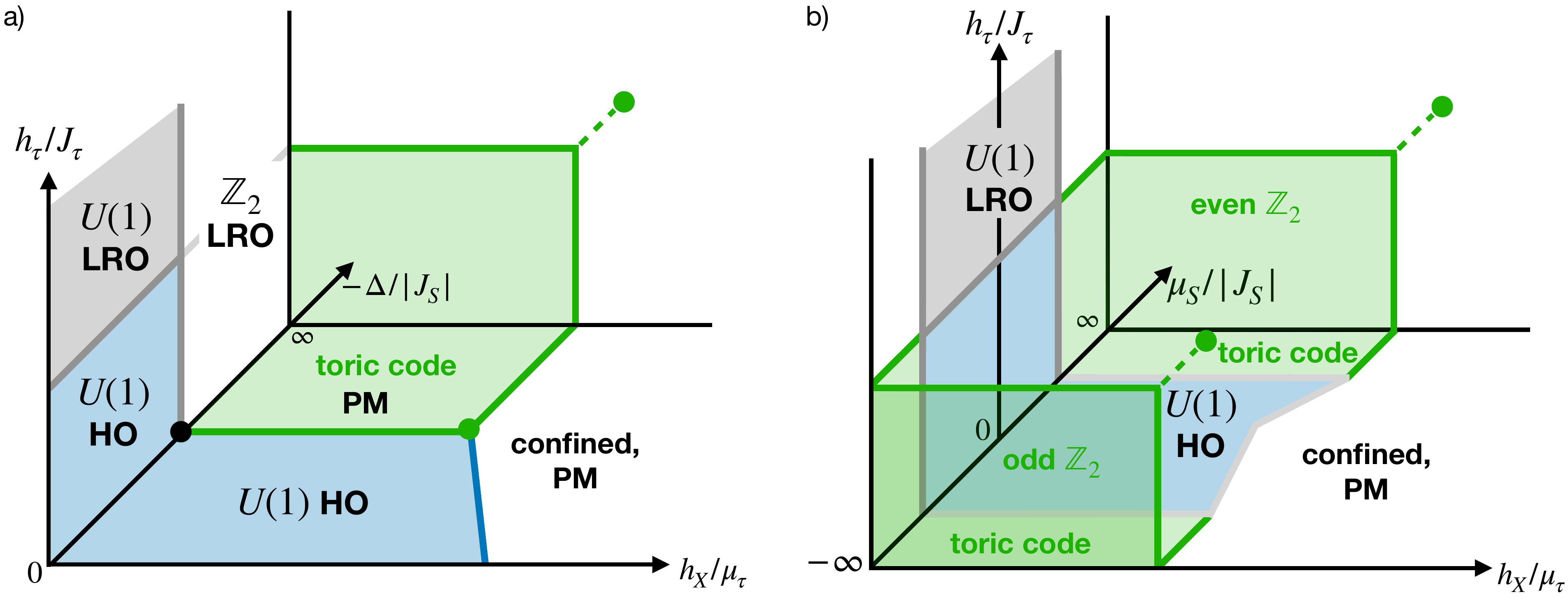}
	\caption{\justifying 
    Schematic phase diagram of the IGT with a $U(1)$ matter field and a $\mathbb{Z}_2$ Higgs field, Eq.~\eqref{eqHU1IGT}. a) In the $S^z_{\rm tot}$ sector corresponding to chemical potential $\mu_S=0$, the HXYO model with $h_X=0$ corresponds to the left vertical plane. Both, the $U(1)$ HO / SPT phase (blue) and the topological $U(1)$ symmetric paramagnet (PM, green), remain stable when open $\tau^z$ strings are introduced by $h_X \neq 0$. For large enough $h_X/\mu_\tau$ the topologically trivial, confined PM is obtained. b) As a function of the chemical potential $\mu_S$ the HXYO model is realized at $h_X=0$ for variable $S^z_{\rm tot}$, with $S^z_{\rm tot}=\pm L^2/2$ for $\mu_S = \mp \infty$ for $\Delta < 0$, where $L^2=\sum_{\vec{j}}$ is the number of lattice sites.
    }
	\label{fig:U1IGT}
\end{figure*}

\emph{Finite-temperature phase diagram.--}
The 2D HXYO model also has a rich finite-temperature phase diagram, see Fig.~\ref{fig:2dHXYO} b). We still consider the closed-loop subspace, $\mu_\tau \to \infty$, and begin our discussion in the regime where $|\Delta| < |J_S|$. Although the long-range $U(1)$ order in squeezed space is immediately destroyed by thermal fluctuations at any $T>0$, below a critical $T^*_{\rm BKT}$ the spins feature quasi-long range order with power-law correlations in squeezed space, $C^*(\vec{i}-\vec{j}) = \langle \hat{S}^+_{\vec{i}} \hat{S}^-_{\vec{j}} \rangle_{\rm sq} \simeq |\vec{i} - \vec{j}|^{-\alpha}$. These correspond to power-law, non-local string correlations in the original basis, where the bare two-point correlator $C(\vec{i}-\vec{j}) = \langle \hat{S}^+_{\vec{i}} \hat{S}^-_{\vec{j}} \rangle \simeq \exp ( |\vec{i} - \vec{j}|/\xi )$ decays exponentially. I.e., as in the 2D HIO model we obtain a finite-temperature SPT phase, which we refer to as the hidden quasi-XY ordered phase, with hidden quasi-long range order. 

Thermodynamically, the finite-$T$ SPT transition out of the hidden-XY ordered phase is in the BKT universality class. At high $T$ a phase with exponential correlations $C^*(\vec{i}-\vec{j}) \simeq \exp ( |\vec{i} - \vec{j}|/\xi^*(T))$ in squeezed space is realized, and similar but with a different correlation length $\xi(T)$ in the original basis. Probing this SPT transition in the BKT class directly in the bulk is likely challenging, due to the infinite order of non-analyticities in thermodynamic properties and the non-local nature of the non-trivial bulk correlations. Hence the most robust probe of the BKT-SPT transition, we believe, is through edge correlations in a system without strings exiting the bulk. As in the 2D HIO phase, in the closed-loop subspace the latter provide direct access to long-range correlations in squeezed space, see discussion around Eq.~\eqref{eqCstring2D}. In the 2D HXYO model we obtain a sequence from long-range, at $T=0$, to power-law, between $0 < T < T^*_{\rm BKT}$, to exponential, for $T>T^*_{\rm BKT}$, edge correlations.

At $|\Delta| = |J_S|$, the XXZ model in squeezed space becomes $SU(2)$ invariant. From the point of view of the original model, Eq.~\eqref{eq:U1_meets_tc}, this is a hidden $SU(2)$ symmetry: constructing it requires the highly non-local unitary $\hat{U}$ to obtain the decoupled XXZ model in squeezed space from which this symmetry is apparent. As a direct consequence of this hidden $SU(2)$ symmetry, the BKT transition disappears at $|\Delta| = |J_S|$. For antiferromagnetic (AFM) interactions, $\Delta = |J_S|>0$, any non-zero $T>0$ leads to a symmetric phase characterized by a non-linear sigma model in squeezed space.

For $\Delta > |J_S|$, i.e. an for easy-axis AFM, the gapped phase retains $\mathbb{Z}_2$, or Ising, long-range order up to a critical temperature $T_{\rm Ising}$ when $S^z_{\rm tot}=0$. For other values of $S^z_{\rm tot}$, the $\mathbb{Z}_2$ symmetry is explicitly broken, the transition disappears and one symmetric, PM phase is formed. The situation for ferromagnetic coupling $\Delta <0$ is similar. As in the ground state, this phase is independent of the loop gas properties.

\emph{Open strings \& Ising gauge theory.--}
Finally, we discuss the effects of open $\hat\tau^z$ strings, introduced by adding a term $\propto h_X$ as in Eq.~\eqref{eqAddhX} to the Hamiltonian Eq.~\eqref{eq:U1_meets_tc}. Again we find it useful to discuss this case in the language of IGT. To this end, we express spins as hard-core bosons via $\hat{S}^+_{\vec{j}}=\hat{a}^\dagger_{\vec{j}}$ ($\hat{S}^-_{\vec{j}}=\hat{a}_{\vec{j}}$) and $\hat{S}^z_{\vec{j}}=\hat{n}^a_{\vec{j}}-1/2$, where $\hat{n}^a_{\vec{j}} = \hat{a}^\dagger_{\vec{j}} \hat{a}_{\vec{j}}$; these bosons $\hat{a}_{\vec{j}}$ correspond to a $U(1)$ matter field. Moreover, we introduce a Higgs field $\hat{\sigma}^x_{\vec{j}}$ on the sites $\vec{j}$ in order to impose the $\mathbb{Z}_2$ Gauss law
\begin{equation}
    (-1)^{\hat{n}^a_{\vec{j}}} \hat{\sigma}^x_{\vec{j}} \prod_{l \in +_{\vec{j}}} \hat{\tau}^x_l \ket{\psi} \equiv \hat{G}_{\vec{j}} \ket{\psi} = \ket{\psi}, \quad \forall ~ \vec{j}.
\end{equation}
This leads to the $U(1)$-plus-IGT Hamiltonian
\begin{multline}
    \hat{H} = - \frac{J_S}{2} \sum_{\nn} \left( \hat{a}^\dagger_{\vec{i}} \hat{\tau}^z_{\nn} \hat{a}_{\vec{j}}  + \hc \right) + \mu_S \sum_{\vec{j}} \hat{n}^a_{\vec{j}}\\
     - h_\tau \sum_{\nn} \hat{\sigma}^z_{\vec{i}} \hat{\tau}^z_{\nn} \hat{\sigma}^z_{\vec{j}}
    - \mu_\tau \sum_{\square} \prod_{l \in \square} \hat \tau^z_l  - \frac{J_\tau}{2} \sum_{\vec{j}} \hat{\sigma}^x_{\vec{j}} \\
     + h_X \sum_{\nn} \hat{\tau}^x_{\nn}+ \Delta \sum_{\nn} (\hat{n}^a_{\vec{i}}-1/2) (\hat{n}^a_{\vec{j}}-1/2) \;,
    \label{eqHU1IGT}
\end{multline}
which commutes with $\hat{G}_{\vec{j}}$.

In Fig.~\ref{fig:U1IGT} a) we show the phase diagram of Eq.~\eqref{eqHU1IGT} at $\mu_S=0$ and assuming ferromagnetic $\Delta < 0$. The vertical plane with $h_X=0$ corresponds to the $T=0$ phase diagram of the HXYO model discussed above. For large $|\Delta| / |J_S| \to \infty$, spins and links factorize in the original basis and the topological toric code phase is known to extend to non-zero values of $h_X/\mu_\tau$. In the horizontal plane, for $h_\tau=0$, the ferromagnetic state turns into a trivial PM because the term $\propto J_\tau$, through the Gauss law, favors a definite orientation of spins $\hat{S}^z$. Since the state is gapped, this phase survives upon introducing $h_X$ and the topologically ordered regime extends into the plane. Likewise, the application of $h_X$-terms creates a gapped excitation, with energy $\propto \mu_\tau$, when starting from the hidden-XY ordered SPT phase. Hence we expect the latter to remain stable upon increasing $h_X$. Eventually, for large enough values of $h_X/\mu_\tau$, open strings with $B_\square=-1$ proliferate and a confined, PM phase is realized. The phase diagram in the $h_\tau=0$ plane resembles that of the $\mathbb{Z}_2$ IGT coupled to soft-core bosons at unit-filling~\cite{Sachdev2019}.

In Fig.~\ref{fig:U1IGT} b) we tune the chemical potential $\mu_S$ and keep $\Delta < 0$ fixed (ferromagnetic coupling). When $\mu_S \to \pm \infty$, trivial polarized states are realized with $S^z_{\rm tot} = \mp {\rm max}$. Comparison to Eq.~\eqref{eq:U1_meets_tc} shows that these states lead to even and odd $\mathbb{Z}_2$ toric codes, respectively, for $J_\tau>0$. In squeezed space, both topological phases are described by the same toric-code Hamiltonian. For intermediate values of $\mu_S$, the gapped, topologically ordered states are connected through the gapless, $U(1)$ HO-SPT phase when $h_X=0$. As in Fig~\ref{fig:U1IGT} a), we conjecture that the HO phase initially remains stable upon increasing $h_X$. For large $h_X$, the trivial, confined PM is realized, which adiabatically connects the trivially polarized spin states. 

Finally, we note that the fate of the gapless $U(1)$ HO-SPT phase at intermediate $h_X$ remains to be analyzed more carefully. In particular, it has been proposed that non-zero $h_X$ can act as a confining force for $\hat{a}$ particles, leading to the formation of $\mathbb{Z}_2$ neutral bosonic pairs whose number $N_{\rm pair} = N_a/2$ is conserved~\cite{Homeier2023}. If the latter condense, a topologically ordered loop gas may remain stable. Whether such a state exists, and how it connects to the $U(1)$ HO phase proposed here for small $h_X$ remains to be worked out.

% % % % % % % % % % % % % % % % % % % % % % % % % % % % 
\section*{Discussion}
\label{sec:Discussion}
% % % % % % % % % % % % % % % % % % % % % % % % % % % % 
In this article we discussed a class of hidden-order symmetry-protected topological (HO-SPT) phases, in which a global symmetry is spontaneously broken but long-range correlations, or any other local order parameter, are hidden by the proliferation of domain walls of the order parameter bound to the strings constituting a loop gas model. Some of the phenomenology we propose has previously been explored, mostly in the context of (emergent) higher-form symmetries~\cite{Verresen2024,Xu2025} and at $T=0$. We went beyond existing studies in three key ways. (i) 
We demonstrate that the HO-SPT phases protected by the closed loop property of the loop-gas model (corresponding to a 1-form symmetry) are stable to thermal fluctuations, in two or more spatial dimensions, leading to rich finite-$T$ phase diagrams. (ii) We construct an exact unitary transformation, which allows to exactly decouple and solve the full spectrum of an entire class of quantum spin models coupled to a loop gas / perturbed toric code. Specifically, we apply this approach to an XY model of spins featuring a continuous $U(1)$ symmetry, which leads to an intrinsically gapless HO-SPT phase with finite-$T$ power-law correlations along appropriate edges and a HO-SPT phase transition in the hidden-BKT universality class. 
(iii) While the non-trivial edge correlations are expected to disappear immediately when open strings of the loop gas are introduced (i.e. when the 1-form symmetry is broken), we put forward a characterization of the $T=0$ and $T>0$ SPT phases in terms of a spontaneously broken global symmetry acting on the spins (which correspond to the matter sector of the equivalent bulk gauge-theory description). Since the mechanism we propose, where SSB of the global spin symmetry is hidden by the fluctuating loop gas, does not require higher-form or local gauge symmetries to protect the SPT phase, we argue that the HO-SPT phases are stable against general perturbations of the loop gas model. Currently we are not aware of an order parameter which is able to detect the hidden SSB without the 1-form symmetry in place, and finding one remains a interesting topic of future research.

Our work has important implications for material sciences: we suggest to search for quantum spin liquids in the HO-SPT class. These states of matter are distinct from the deconfinded, topological toric code phase that underlies several gapped, $\mathbb{Z}_2$ topological quantum spin liquids~\cite{Savary2017,Rokhsar1988,Verresen2021,Samajdar2021,Semeghini2021,Wang2025}, but also feature short-range correlations without local order parameters in the bulk. Such HO-SPT phases might most easily be identified  through their unconventional effect on edge correlations, and they are robust at finite temperatures. Another system in which hidden order has recently been proposed to play a role by some of us is the pseudogap (PG) phase of hole-doped cuprate superconductors~\cite{Schloemer2024}. Specifically, we suggested that a fluctuating string net of AFM stripes may hide an underlying broken $SU(2)$ symmetry, leading to a fractionalized orthogonal metal. Building upon this picture, we speculate that the finite-temperature transition into the PG phase, at a characteristic temperature $T^*$ that depends on the hole doping, might realize a finite-$T$ HO-SPT transition protected by the global $SU(2)$ symmetry~\cite{Paciani2025} ending in a hQCP around optimal doping, where the $SU(2)$ symmetry is fully restored. Exploring the physics of $SU(2)$ HO-SPT phases will be an important future task enabled by our work. 

While we laid out the basic physics of HO-SPT phases, understanding further signatures of HO-SPT phases constitutes an important future research direction~\cite{Wilke20252}. For example, for systems with hidden continuous symmetry breaking we predict gapless Goldstone modes in the spectrum. Since the latter involve strong entanglement between links and spins, their spectral signatures in the HO-SPT phase remain to be clarified. It will also be interesting to study the relation between HO-SPT phases and nematic states of matter, which can also be described in the framework of lattice gauge theories~\cite{Podolsky2005}.

Finally, we propose to search for HO-SPT phases directly in quantum simulators. One avenue is to start from digital schemes, where the perturbed toric code has already been successfully implemented~\cite{Cochran2024}. The stability of the HO-SPT phase to thermal fluctuations also suggests some intrinsic robustness to noise, and the study of edge-correlations in mixed quantum states constitutes a promising route to demonstrate this effect. On the other hand, Rydberg tweezer arrays~\cite{Bernien2017} constitute a promising alternative platform for directly realizing HO-SPT phases. Specifically, by using so-called local pseudogenerators~\cite{Halimeh2022}, it is possible to implement $\mathbb{Z}_2$ lattice gauge theories on the honeycomb lattice with couplings to $\mathbb{Z}_2$ or $U(1)$ matter fields~\cite{Homeier2023}, and implement the toy models introduced in this work: the HIO model in 1D or 2D, and the HXYO model.

\section*{Acknowledgments}
The authors acknowledge inspiring discussions with G. De Paciani, G. D\"unnweber, L. Homeier, T. Rakovszky, R. Verresen, M. Kebric, S. Paeckel, S. Gazit, E. Demler and L. Pollet. This project has received funding from the European Research Council (ERC) under the European Union’s Horizon 2020 research and innovation programm (Grant Agreement no 948141) — ERC Starting Grant SimUcQuam, and by the Deutsche Forschungsgemeinschaft (DFG, German Research Foundation) under Germany's Excellence Strategy -- EXC-2111 -- 390814868.

% % % % % % % % % % % % % % % % % % % % % % % % % % % % 
%\section*{Methods}
%\label{secMethods}
% % % % % % % % % % % % % % % % % % % % % % % % % % % % 
\appendix
\section{Hidden order transformation}
\label{sec:appA}

\emph{1D HIO model.--}
We explicitly construct the unitary transformation $\hat{U}$ disentangling link and spin degrees of freedom at $\lambda = 0$. 
In the main text we defined $\hat{U}$ by its action on the basis states, see Eq.~\eqref{eq:1d_HO_unitary}.
The explicit operator form of the unitary is given by
\begin{equation}
    \hat U = \prod_j \hat U_j = \prod_j (2 \hat S^x_j)^{ \hat{\gamma}_j},
    \quad \hat{\gamma}_j = \frac{1}{2}(1-\hat{\pi}_j)\;,
\end{equation}
with the parity $\hat{\pi}_{j}=(-1)^{\hat{p}_{j}} = \prod_{i<j} \hat{\tau}^z_{\langle i, i+1\rangle}$.
Furthermore, it holds that $\hat U^\dagger = \hat U$ and $\hat U^2 = \mathbb{1}$.
As a consequence, a given spin on site $j$ is flipped if the number of links $\tau^z_{\langle i,i+1 \rangle}=-1$ with $i<j$ is odd.
Let us consider the action of $\hat U$ on~\eqref{eq:1d_ham}, in particular the terms $\propto J_S, J_\tau$ in Eq.~\ref{eq:1d_ham} on which it acts non-trivially.
The term $\propto J_S$ transforms as
\begin{equation}
    \hat U^\dagger \hat S^z_j \hat S^z_{j+1} \hat \tau^z_{\langle j, j+1\rangle} \hat U = \hat S^z_j \hat S^z_{j+1} (\hat \tau^z_{\langle j, j+1\rangle})^2 = \hat S^z_j \hat S^z_{j+1}\;,
\end{equation}
where we made use of the anti-commutation relation of spins $\{ \hat{S}^\alpha, \hat{S}^\beta \} = \frac{1}{2}\delta_{\alpha\beta}$ and $ \hat{\gamma}_{j+1} = \hat{\gamma}_j +1 ~{\rm mod}~2$ ($ \hat{\gamma}_{j+1} = \hat{\gamma}_j ~{\rm mod}~2$) if $ \tau^z_{\langle j, j+1\rangle} = -1$ $(\tau^z_{\langle j, j+1\rangle} = 1)$.
From $\hat \tau^x_{\langle j, j+1\rangle} (2\hat S^x_i)^{\hat \gamma_i} = (2\hat S^x_i)^{\hat \gamma_i +1} \hat \tau^x_{\langle j, j+1\rangle}$ for $i\geq j+1$, it follows 
\begin{equation}
    \hat U^\dagger \hat \tau^x_{\langle j, j+1\rangle}\hat U = \prod_{i \geq j+1} (2\hat S^x_i) \hat \tau^x_{\langle j,j+1\rangle}\;,
\end{equation}
i.e. the unitary attaches a string of $\hat S^x_i$ to $\hat \tau^x_{\langle j, j+1\rangle}$.
As a result, the term $\propto J_\tau$ transforms as
\begin{equation}
\begin{aligned}
    \hat U^\dagger \hat \tau^x_{\langle j-1, j\rangle} \hat \tau^x_{\langle j, j+1\rangle} \hat S^x_j \hat U &= (2 \hat S^x_j) \hat S^x_j \hat \tau^x_{\langle j-1,j\rangle}\hat \tau^x_{\langle j,j+1\rangle}\\ &= \frac{1}{2} \hat \tau^x_{\langle j-1,j\rangle}\hat \tau^x_{\langle j,j+1\rangle}\;.
\end{aligned}
\label{eqUtransform1DHIO}
\end{equation}
The final decoupled Hamiltonian is given by
\begin{equation}
\begin{aligned}
	\hat U^\dagger \hat H \hat U&= - J_S \sum_{j} \hat S_{j+1}^z \hat S_j^z
    + h_S \sum_j \hat S^x_j\\
    &- h_\tau \sum_j \hat \tau^z_{\langle j,j+1\rangle}
    + \frac{J_\tau}{2} \sum_j \hat \tau^x_{\langle j-1,j\rangle} \hat\tau^x_{\langle j,j+1\rangle}\\
    &= \hat H^S_{\rm TFIM} + \hat H^\tau_{\rm TFIM}
    \;.
\end{aligned}
\end{equation}

\emph{2D HIO model.--}
In a similar manner, we demonstrate the action of the unitary transformation $\hat U$ in the $2D$ HIO model:
\begin{equation}
    \hat U = \prod \hat U_\mathbf{j} = \prod_\mathbf{j} (2 \hat S^x_\mathbf{j})^{ \hat \gamma_\mathbf{j}},
    \quad \hat\gamma_\mathbf{j} = \frac{1}{2}(1-\hat \pi_\mathbf{j})\;.
\end{equation}
In 2D, the  parity is defined as $(-1)^{\hat{p}_{\mathbf{j}}}=\prod_{l\in \mathcal{L}_{\mathbf{j}}} \hat \tau_l^z$, where $\mathcal{L}_{\mathbf{j}}$ denotes a path on the links starting at a fixed reference site and leading to $\mathbf{j}$. While there is only one choice of trajectory $\mathcal{L}_{\mathbf{j}}$ in the 1D model, care has to be taken in the 2D model. 

For the first term in Eq.~\eqref{eq:TC_meets_Ising} that transform non-trivially under $\hat U$, the one $\propto J_S$, any choice of $\mathcal{L}_{\vec{j}}$ starting at the same, fixed reference site for all $\vec{j}$ leads to the following transformation:
\begin{equation}
\begin{aligned}
    &\hat U^\dagger \hat S^z_\mathbf{i} \hat S^z_\mathbf{j} \hat \tau^z_{\langle \mathbf{i},\mathbf{j} \rangle} \hat{U}\\ &= 
    \begin{cases}
        (\pm 1)^2 \hat \tau^z_{\langle \mathbf{i},\mathbf{j} \rangle} \hat S_\mathbf{i}^z \hat S_\mathbf{j}^z\;, \quad &\tau^z_{\langle\mathbf{i},\mathbf{j}\rangle} = 1\;\\
        - \hat \tau^z_{\langle \mathbf{i},\mathbf{j} \rangle} \hat S_\mathbf{i}^z \hat S_\mathbf{j}^z\;, \quad &\tau^z_{\langle \mathbf{i},\mathbf{j} \rangle} = - 1\; 
    \end{cases}\\
    &=(\hat{\tau}^z_{\langle \mathbf{i},\mathbf{j} \rangle})^2 \hat S_\mathbf{i}^z \hat S_\mathbf{j}^z = \hat S_\mathbf{i}^z \hat S_\mathbf{j}^z\;,
\end{aligned}
\end{equation}
since we work in the closed loop subspace.

\begin{figure}[t]
    \centering
	\includegraphics[width=.25\textwidth]{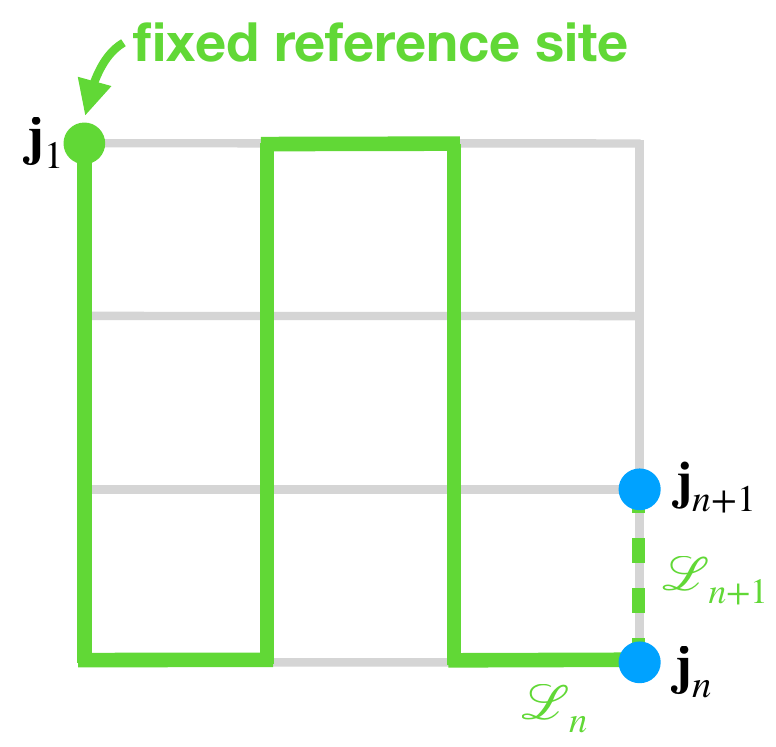}
	\caption{\justifying 
    Hidden order transformation $\hat{U}$ for the 2D HIO and 2D HXYO models. We choose a particular set of paths $\mathcal{L}_{\vec{j}_n}$, such that for a consecutive set of nearest-neighbor sites $\langle \vec{j}_{n+1}, \vec{j}_n \rangle$, the trajectory $\mathcal{L}_{\vec{j}_{n+1}}$ is obtained from $\mathcal{L}_{\vec{j}_n}$ by adding the nearest-neighbor link $\langle \vec{j}_{n+1}, \vec{j}_n \rangle$. Site $\vec{j}_1$ where all $\mathcal{L}_{\vec{j}_n}$ start defines a fixed reference site.
    }
	\label{fig:snake}
\end{figure}

To decouple spins from links in the second term, $\propto J_\tau$, a particular choice of trajectories $\mathcal{L}_{\vec{j}}$ is required. As illustrated in Fig.~\ref{fig:snake}, we consider a recursive construction of $\mathcal{L}_{\vec{j}}$: it starts by labeling all sites by a one-dimensional snake $\vec{j}_n$, with $n=1...N_s$ where $N_s$ is the total number of sites of the lattice; $\vec{j}_1$ is the reference sites common to all $\mathcal{L}_n \equiv \mathcal{L}_{\vec{j}_n}$. The snake is chosen such that any set of consecutive sites constitutes a nearest-neighbor pair, $\langle \vec{j}_{n+1}, \vec{j}_n \rangle$ for all $n$. The trajectory $\mathcal{L}_1$ is trivial and contains only the reference site $\vec{j}_1$. In every step of the recursion, $\mathcal{L}_{n+1}$ is obtained from $\mathcal{L}_n$ by starting from $\mathcal{L}_n$ and extending it by the link $\langle \vec{j}_{n+1}, \vec{j} \rangle$. Thereby, every trajectory $\mathcal{L}_{m} \supset \mathcal{L}_n$ for $m>n$ contains all previous trajectories. 

Now we apply the unitary transformation $\hat{U}$ to the term $\propto J_\tau$ in Eq.~\eqref{eq:TC_meets_Ising}. Since $\hat{\tau}^z_{\nn}$ on bonds $\nn$ which are not part of the snake do not appear in any $\mathcal{L}_{\vec{r}}$, these variables commute with $\hat{U}$ and can be viewed as $c$-numbers in the following. Hence, $\hat{\tau}^z_{\langle \vec{j}_{n+1},\vec{j}_n\rangle}$ on bonds $\langle \vec{j}_{n+1},\vec{j}_n\rangle$ which are part of the snake form an effective 1D HIO model, along the snake, and we obtain the same transformation law as above, see Eq.~\eqref{eqUtransform1DHIO}. Making use of $(2\hat S^x_j)^{\hat\gamma_j}\hat \tau^x_{\langle j-1, j\rangle} (2\hat S^x_j)^{\hat\gamma_j} = (2\hat S^x_j)^{\hat\gamma_j}(2\hat S^x_j)^{\hat \gamma_j + 1}\hat \tau^x_{\langle j-1, j\rangle} = 2 \hat S^x_j \hat \tau^x_{\langle j-1, j\rangle}$, the term $\propto J_\tau$ transforms as 
\begin{equation}
\begin{aligned}
    \hat U^\dagger \left( \prod_{l \in +_\mathbf{j}} \hat \tau^x_l \right) \hat S^x_\mathbf{j} \hat{U} 
    &= (2\hat S_\mathbf{j}^x)^{\hat\gamma_\mathbf{j}} \left[ \left( \prod_{l \in +_\mathbf{j}} \hat \tau^x_l \right) \hat S^x_\mathbf{j} \right](2 \hat S_\mathbf{j}^x)^{\hat\gamma_\mathbf{j}} \\&= (2\hat S_\mathbf{j}^x)\hat S^x_\mathbf{j}\prod_{l \in +_\mathbf{j} } \hat \tau^x_l\ = \frac{1}{2} \prod_{l \in +_\mathbf{j} } \hat \tau^x_l\;.
\end{aligned}
\end{equation}
Hence, the unitary transformation allows us to decouple the Ising and toric code model,
\begin{equation}
    \hat U^\dagger \hat H \hat U = \hat H_{\rm TC-F} + \hat H_{\rm TFIM}\;.
\end{equation}

\emph{2D HXYO model.--}
In this case, the unitary transformation is defined as 
\begin{equation}
    \hat U = \prod \hat U_\mathbf{j} = \prod_\mathbf{j} (2 \hat S^z_\mathbf{j})^{ \hat \gamma_\mathbf{j}},
    \quad \hat\gamma_\mathbf{j} = \frac{1}{2}(1-\hat \pi_\mathbf{j})\;,
\end{equation}
with the path defined in the same way as for the 2D HIO model.
The calculation can be performed in a similar manner, making use of the anti-commutation of spins and 
\begin{equation}
\begin{aligned}
    &\hat U^\dagger \hat S^\mu_\mathbf{i} \hat S^\mu_\mathbf{j} \hat \tau^z_{\langle \mathbf{i},\mathbf{j} \rangle} \hat{U}\\ &= 
    \begin{cases}
        (\pm 1)^2 \hat \tau^z_{\langle \mathbf{i},\mathbf{j} \rangle} \hat S_\mathbf{i}^\mu \hat S_\mathbf{j}^\mu\;, \quad &\tau^z_{\langle\mathbf{i},\mathbf{j}\rangle} = 1\;\\
        - \hat \tau^z_{\langle \mathbf{i},\mathbf{j} \rangle} \hat S_\mathbf{i}^\mu \hat S_\mathbf{j}^\mu\;, \quad &\tau^z_{\langle \mathbf{i},\mathbf{j} \rangle} = - 1\; 
    \end{cases}\\
    &=(\tau^z_{\langle \mathbf{i},\mathbf{j} \rangle})^2 \hat S_\mathbf{i}^\mu \hat S_\mathbf{j}^\mu = \hat S_\mathbf{i}^\mu \hat S_\mathbf{j}^\mu\;.
\end{aligned}
\end{equation}
with $\mu = x,y$.

\section{Numerical DMRG simulations: 1D HIO model}
\label{sec:appB}
We perform DMRG simulations using the \textsc{SyTen} toolkit to evaluate magnetization $M_S$ ($M_S^*$) in real (squeezed) space, as well as the correlator $|\langle \hat S^z_0 \hat S^z_{x} \rangle|$.
The former are computed from snapshots of the many-body wavefunction via the perfect sampling approach~\cite{Ferris2012,Buser2022}.
The sample-averaged magnetization of the state of a system of $L$ sites is computed as
\begin{equation}
    M = \frac{1}{N} \sum_i |M_i|, \quad M_i = \frac{1}{L} \sum_j  S_{j}^z |_i\;.
\end{equation}
Here, $S_{j}^z|_i$ denotes the spin-$z$ value on site $j$ in snapshot $i$ and $N = 10^4$ denotes the total number of sampled snapshots.
In analogy, the sample-averaged squeezed space magnetization is computed with respect to the squeezed space value $S_{j}^z|_{{\rm sq},i}$, i.e. the unitary~\eqref{eq:1d_HO_unitary} is applied to the Fock basis state prior to the evaluation of $S_{j}^z$.

\bibliography{bibliography}

\end{document}

%% file: header.tex
\usepackage{graphicx}
\usepackage{latexsym}
\usepackage{amssymb}
\usepackage{amsmath}
\usepackage{amsfonts}
\usepackage{enumerate}
\usepackage{upgreek}
\usepackage{subfigure}
\usepackage{bm}
\usepackage{nicefrac}
\usepackage{bbold}
\usepackage{verbatim}
\usepackage{multirow}
\usepackage{color}
\usepackage{comment}
\usepackage{xcolor}
\usepackage{soul}
\usepackage{tikz}
\usepackage{braket}
\usepackage{caption}
\usepackage{ragged2e}

\newcommand{\hc}{\mathrm{H.c.}}
\renewcommand{\vec}[1]{\mathbf{#1}}
\newcommand{\nn}{\langle \mathbf{i}, \mathbf{j}\rangle}